\documentclass[superscriptaddress,preprint]{revtex4}

\usepackage{graphicx}

\begin{document}

\title{Percolation theory applied to measures of fragmentation in social
  networks}

\author{Yiping Chen}

\affiliation{Center for Polymer Studies, Boston University, Boston,
  Massachusetts 02215, USA}

\author{Gerald Paul}

\affiliation{Center for Polymer Studies, Boston University, Boston,
  Massachusetts 02215, USA}

\author{Reuven Cohen} 

\affiliation{Department of Electrical and Computer Engineering, Boston
  University, Boston, Massachusetts 02215, USA}

\author{Shlomo Havlin}

\affiliation{Minerva Center and Department of Physics, Bar-Ilan University,
  52900 Ramat-Gan, Israel}

\author{Stephen P. Borgatti}

\affiliation{Dept of Org. Studies, Boston College, Chestnut Hill, MA 02467,
  USA}

\author{Fredrik Liljeros}

\affiliation{Department of Sociology, Stockholm University, S-106 91
  Stockholm, Sweden}

\author{H. Eugene Stanley}

\affiliation{Center for Polymer Studies, Boston University, Boston,
  Massachusetts 02215, USA}

\date{\today}

\begin{abstract}
  We apply percolation theory to a recently proposed measure of fragmentation
  $F$ for social networks. The measure $F$ is defined as the ratio between
  the number of pairs of nodes that are not connected in the fragmented
  network after removing a fraction $q$ of nodes and the total number of
  pairs in the original fully connected network. We compare $F$ with the
  traditional measure used in percolation theory, $P_{\infty}$, the fraction
  of nodes in the largest cluster relative to the total number of nodes.
  Using both analytical and numerical methods from percolation, we study
  Erd\H{o}s-R\'{e}nyi (ER) and scale-free (SF) networks under various types
  of node removal strategies. The removal strategies are: random removal,
  high degree removal and high betweenness centrality removal. We find that
  for a network obtained after removal (all strategies) of a fraction $q$ of
  nodes above percolation threshold, $P_{\infty}\approx (1-F)^{1/2}$.  For
  fixed $P_{\infty}$ and close to percolation threshold ($q=q_c$), we show
  that $1-F$ better reflects the actual fragmentation.  Close to $q_c$, for a
  given $P_{\infty}$, $1-F$ has a broad distribution and it is thus possible
  to improve the fragmentation of the network. We also study and compare the
  fragmentation measure $F$ and the percolation measure $P_{\infty}$ for a
  real social network of workplaces linked by the households of the employees
  and find similar results.

\end{abstract}

\maketitle

\section{Introduction}

Many physical, sociological and biological systems are represented by complex
networks~\cite{paul,pastor2,doro, bocca, albert, cohen, newman,cohen2, paxon,
  callaway, valente, gerry, chung,burda,song,freeman, wasserman}. One of the
important problems in complex networks is the fragmentation of
networks~\cite{cohen, callaway, cohen2,valente, gerry, chung, burda}. In this
problem one studies the statistical properties of the fragmented networks
after removing nodes (or links) from the original fully connected network
using a certain strategy.  Many different removal strategies have been
developed for various purposes, e.g., mimicking the real world network
failures, improving the effectiveness of network disintegration, etc.
Examples include random removal (RR) strategy, the high degree removal (HDR)
strategy and the high betweenness centrality removal strategy
(HBR)~\cite{valente, tanizawa, pastor,holme, newman0}. Note that the best
strategy for fragmentation (minimum nodes removal) is also the best for
immunization since it represent the minimum number of nodes or links needed to
be immunized so that epidemic cannot spread in the network.

Recently, a new measure of fragmentation has been developed in social network
studies~\cite{borgatti}.  Given a fully connected network of $N$ nodes which
is fragmented into separate clusters~\cite{footnote} by removing $m$ nodes
following a certain strategy. We define $q\equiv m/N$ the concentration of
nodes removed and $p\equiv 1-q$ the concentration of existing nodes. The
degree of fragmentation $F$ of the network is defined as the ratio between
the number of pairs of nodes that are not connected in the fragmented network
and the total number of pairs in the original fully connected network.
Suppose that after removal there are $n$ clusters in the fragmented network,
since all members of a cluster are, by definition, mutually reachable, the
measure $F$ can be written as follows~\cite{borgatti}
\begin{equation}
F\equiv 1-\frac{\Sigma^n _{j=1} N_j(N_j-1)}{N(N-1)}\equiv 1-C.
\label{f_def}
\end{equation}
Here, $N_j$ is the number of nodes in cluster $j$, $n$ is number of clusters
in the fragmented network, and $N$ the number of nodes in the original fully
connected network. For an undamaged network, $F=0$.  For a totally
fragmented network, $F=1$. The quantity $C$ defined in Eq.~(\ref{f_def}) can
be regarded as the ``connectivity'' of the network.  When $C=1$ the network
is fully connected while for $C=0$ it is fully fragmented. 

In this paper, we study the statistical behavior of $F\equiv 1-C$ using both
analytical and numerical methods and relate it to the traditional measure of
fragmentation, the relative size of the largest cluster, $P_{\infty}$, used in
percolation theory. In this way, we are able to obtain analytical results for
the fragmentation $F$ of networks. We study three removal strategies: the
\emph{random removal (RR) strategy} which removes randomly selected nodes,
the \emph{high degree removal (HDR) strategy} which targets and removes nodes
with highest degree and the \emph{high betweenness centrality removal (HBR)
  strategy} which targets and removes nodes with highest betweenness
centrality.  The HDR (or HBR) strategies first removes the node with the
highest degree (or the highest betweenness centrality), and then the second
highest and so on.  These three strategies are commonly used in models
representing random and targeted attacks in real world
networks~\cite{albert,callaway,cohen,cohen2, holme}.

\section{Theory}

Traditionally, in analogy to percolation, physicists describe the
connectivity of a fragmented network by the ratio $P_{\infty}\equiv
N_{\infty}/N$ (called the incipient order parameter) between the largest
cluster size $N_{\infty}$ (called the infinite cluster) and $N$.  Many
properties have been derived for this measure~\cite{dietrich,cohen,11a}. For
example, in random networks, $P_{\infty}$ undergoes a second order phase
transition at a threshold $p_c$.  Below $p_c$, $P_{\infty}$ is zero for $N
\to \infty$, while for $p>p_c$, $P_{\infty}$ is finite. This occurs for both
RR and HDR in random networks and lattice
networks~\cite{cohen,callaway,cohen2,11a, dietrich}.  The threshold parameter
$p_c$ depends on the degree distribution, the network topology, and the
removal strategy~\cite{callaway,cohen,cohen2,dietrich,11a}.  The specific way
that $P_{\infty}$ approaches zero at $p_c$ depends on the network topology
and removal strategy but not on details such as $p_c$.  In scale free
networks, where the degree distribution $p(k)\sim k^{-\lambda}$ and
$2<\lambda<3$, it has been found that $p_c\to 0$ for RR strategy~\cite{cohen}
while $p_c$ is very high for HDR strategy~\cite{callaway, cohen2} and for HBR
strategy~\cite{holme}.  For
$\lambda>3$ and RR, $p_c$ is finite.

\begin{figure}[!ht]
\includegraphics[width=0.45\textwidth]{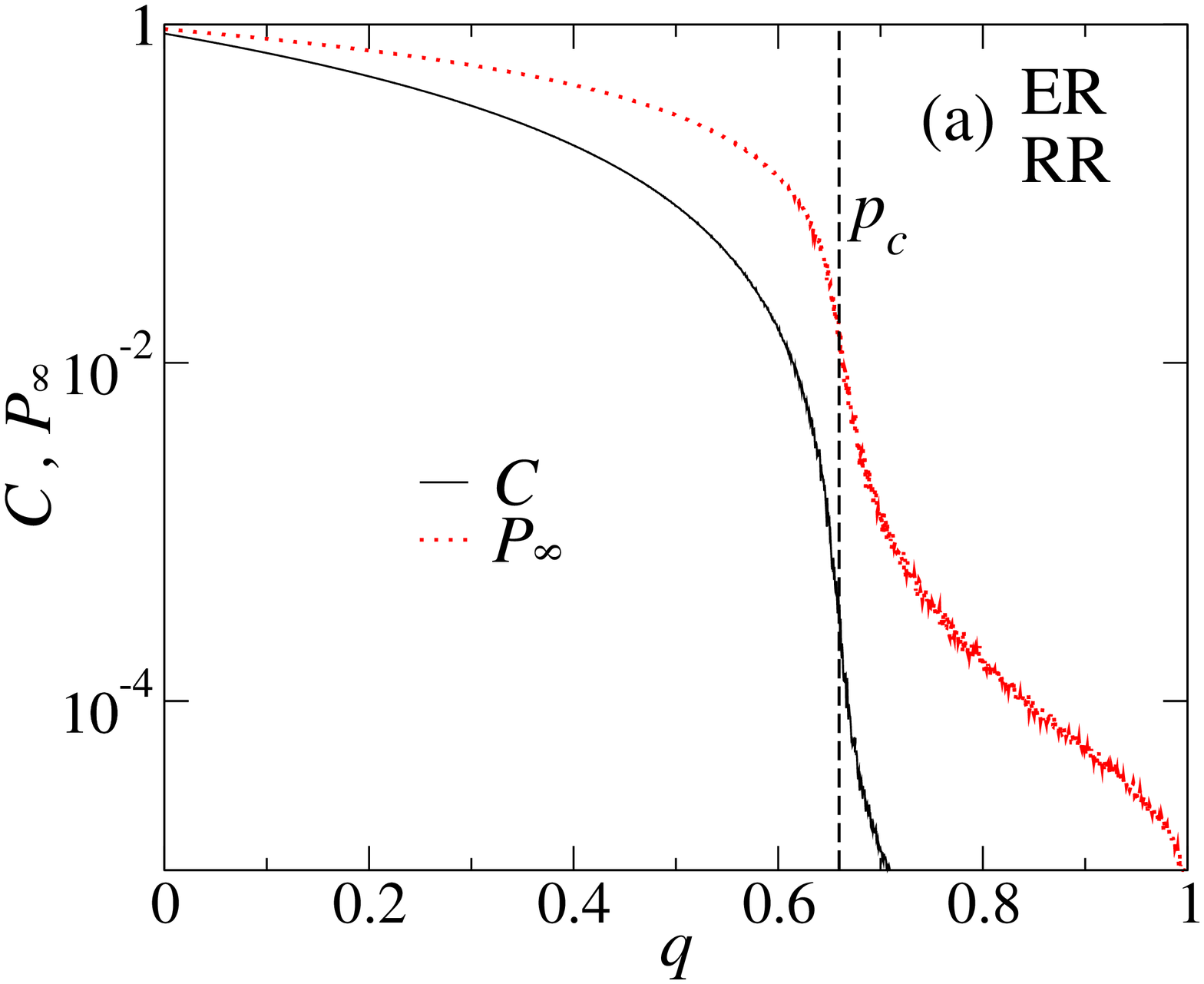}
\includegraphics[width=0.45\textwidth]{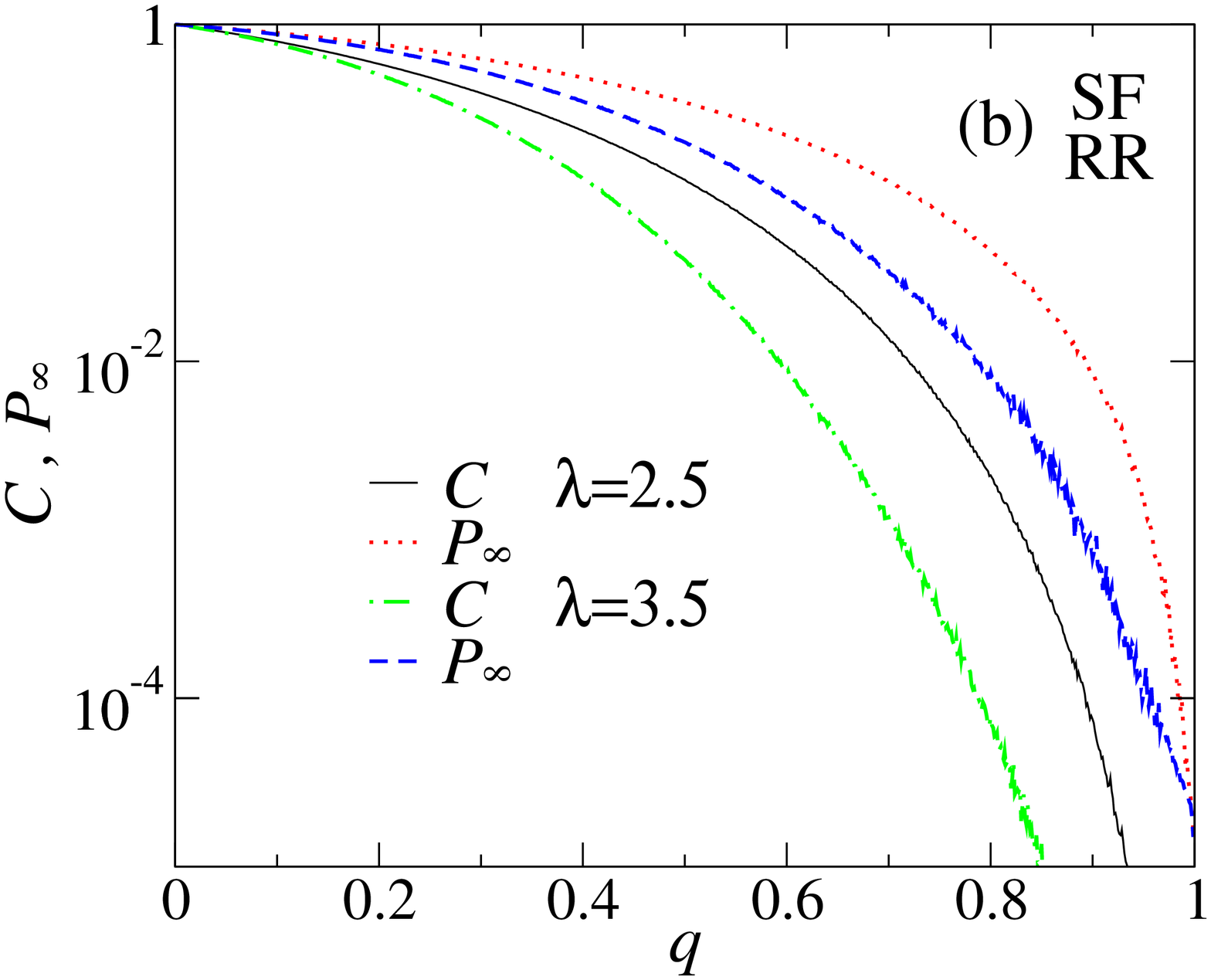}
\includegraphics[width=0.45\textwidth]{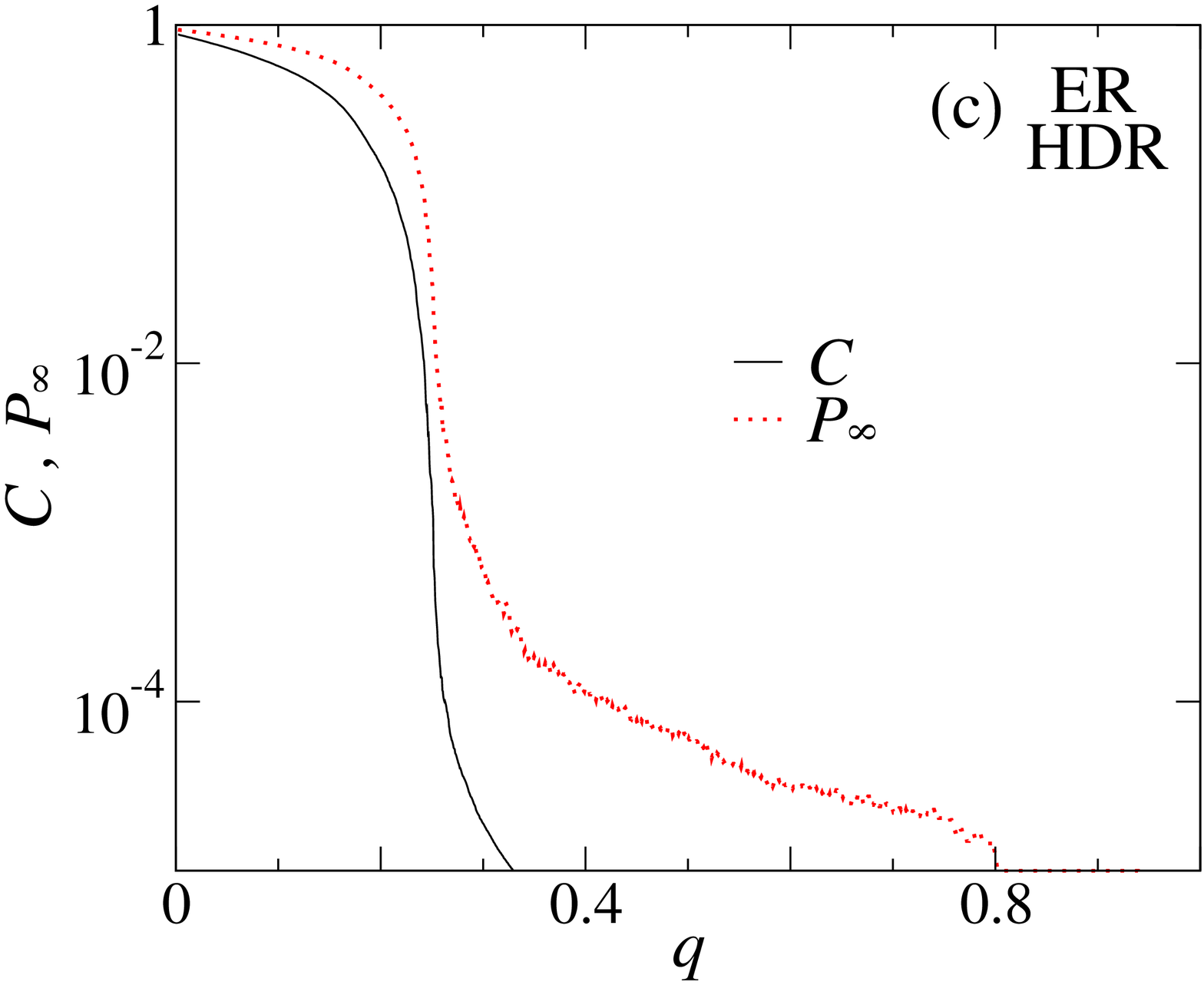}
\includegraphics[width=0.45\textwidth]{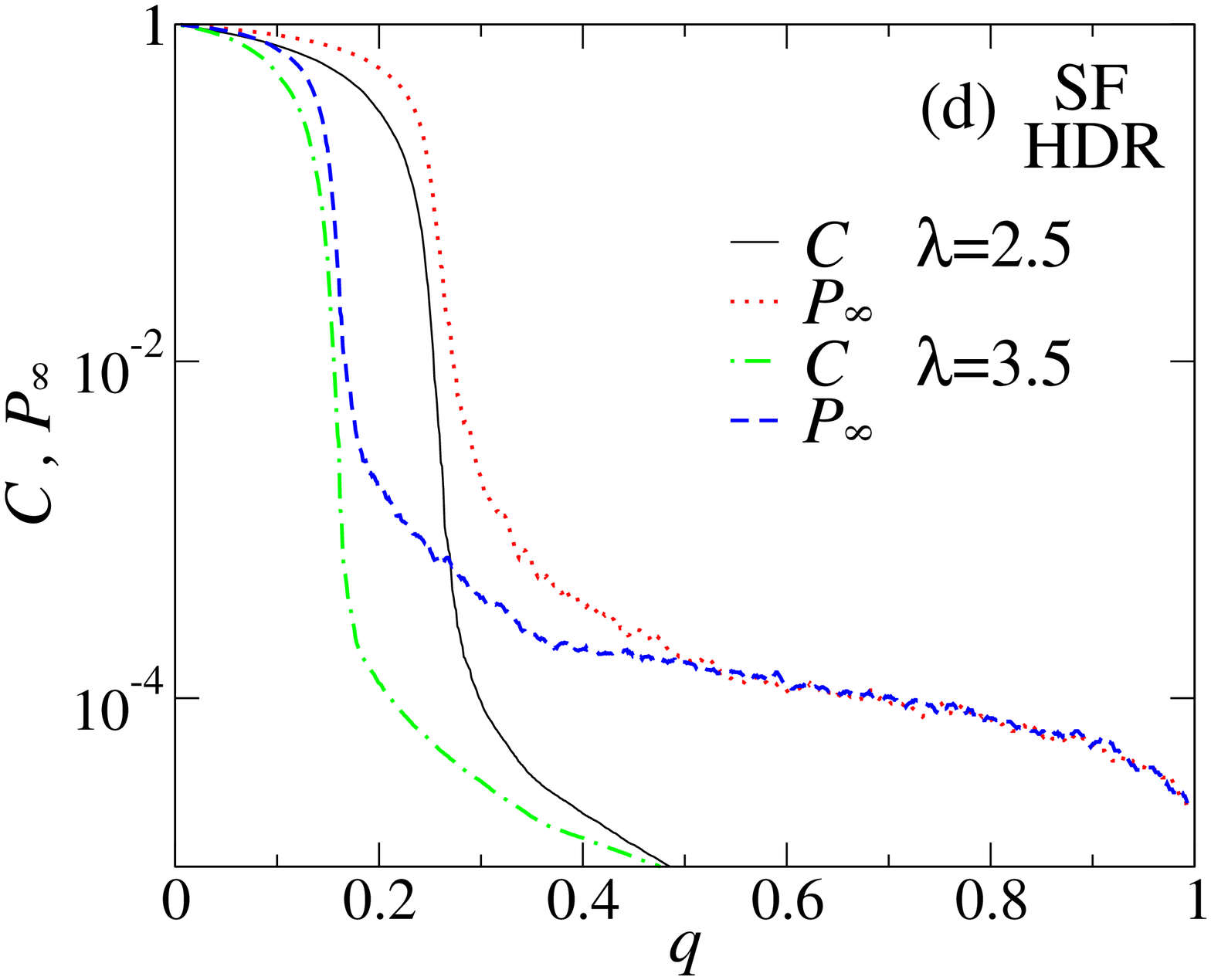}
\includegraphics[width=0.45\textwidth]{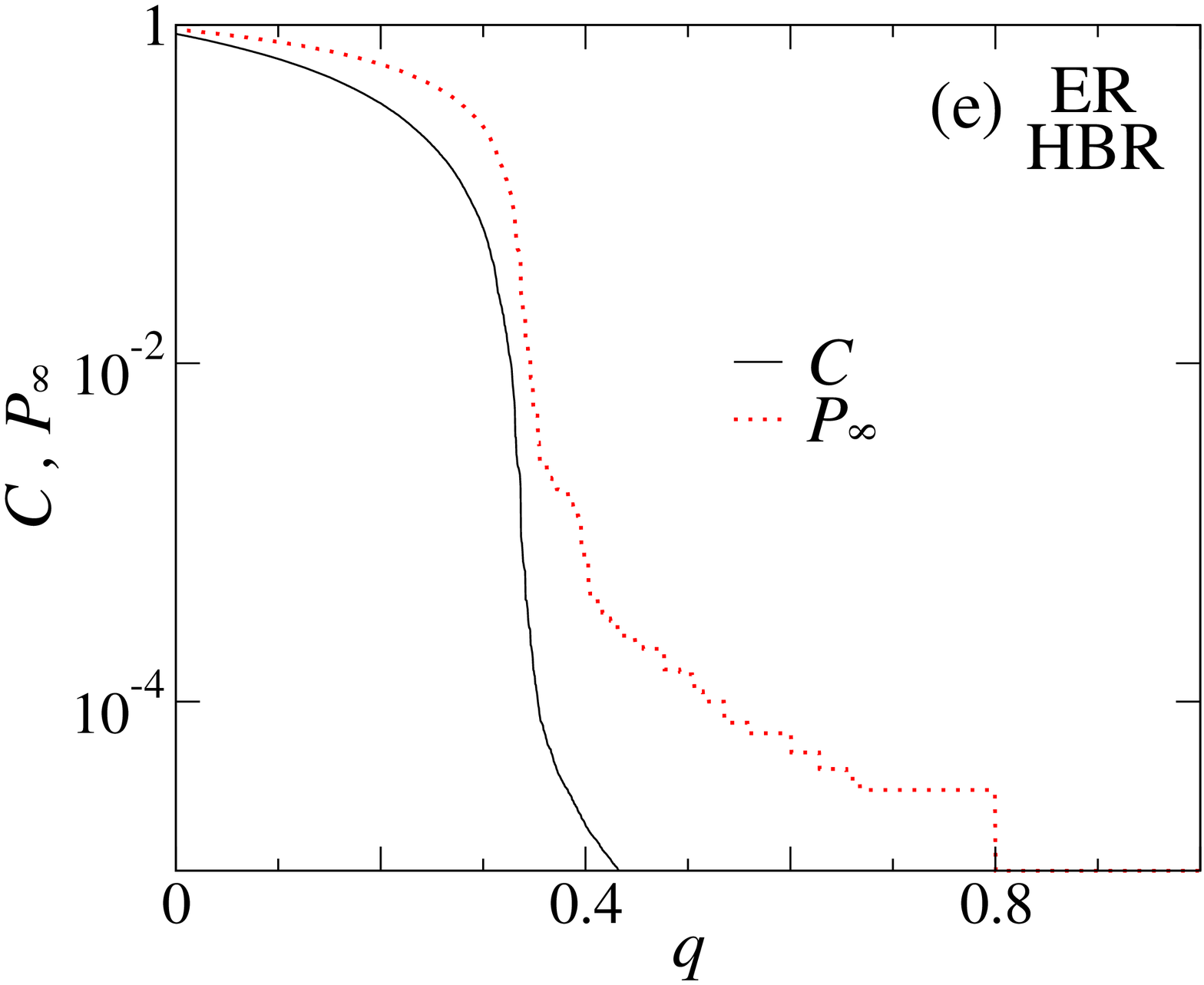}
\includegraphics[width=0.45\textwidth]{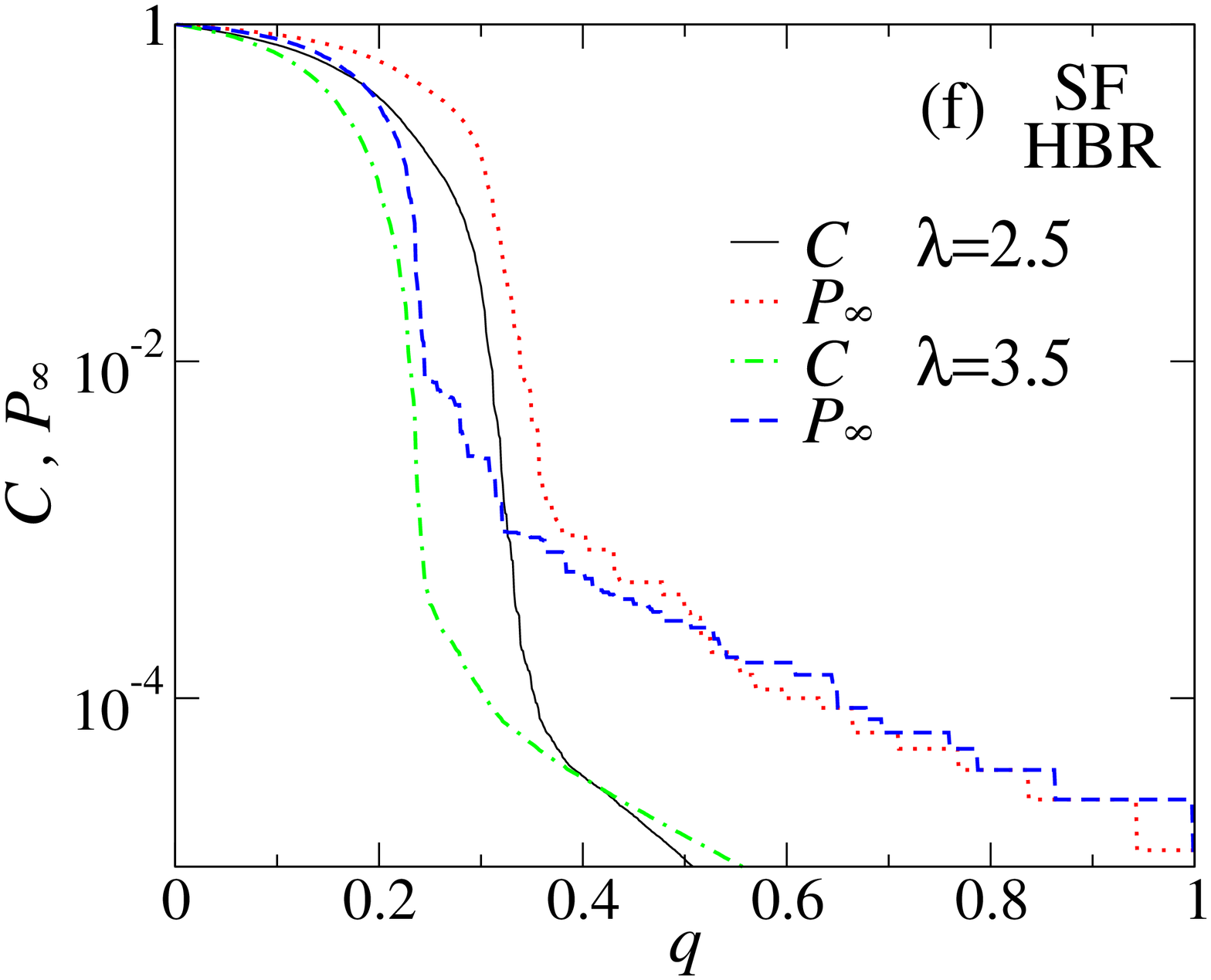}
\caption{The behavior of $C$ and $P_{\infty}$ versus $q$ on ER and SF
  networks. For ER networks, $N=200000$ and $\langle k\rangle=3$. For SF
  networks, $N=80000$. The graphs are (a) RR strategy on ER networks, (b) RR
  strategy on SF networks, (c) HDR strategy on ER networks, (d) HDR
  strategy on SF networks, (e) HBR strategy on ER networks and (f) HBR
  strategy on SF networks.\label{f_kill}}
\end{figure}

Next, we show simulation results of removing nodes in all strategies (RR, HDR
and HBR) on ER and scale free networks. Fig.~\ref{f_kill} shows the behavior
of $C$ ($\equiv 1-F$) and $P_{\infty}$ versus $q$ for Erd\H{o}s-R\'{e}nyi
(ER) and scale-free (SF) networks with RR (Fig.~\ref{f_kill}(a),(b)), HDR
(Fig.~\ref{f_kill}(c),(d)) and HBR (Fig.~\ref{f_kill}(e),(f)) strategies. As
seen in Fig.~\ref{f_kill}(a), the network becomes more fragmented when $q$
increases and both measures drop sharply at $q_c=1-p_c$. Note that $C$ shows
a transition similar to $P_{\infty}$ at $p=p_c$; however, above $q_c$, $C$
becomes more flat in contrast to $P_{\infty}$, indicating the effect of
connectivity in the small clusters which do not effect $P_{\infty}$.

In contrast to Fig.~\ref{f_kill}(a), the transition in Fig.~\ref{f_kill}(b)
is not as sharp and therefore $C$ and $P_{\infty}$ do not show a collapse
together. The reason is that for $\lambda=2.5$ there is no transition at
$q<1$~\cite{cohen} and for $\lambda=3.5$, $P_{\infty}$ falls much less
sharply compared to ER~\cite{cohen3}. For HDR shown in
Figs.~\ref{f_kill}(c),(d), the transition is again sharp since after removing
high degree nodes, the network becomes similar to ER networks, which do not
have high degree nodes~\cite{cohen2}. A similar behavior is seen for HBR
shown in Figs.~\ref{f_kill}(e),(f) due to the known high correlation between
high degree nodes and high betweenness centrality nodes~\cite{holme}.

When $p> p_c$ and not too close to $p_c$, following percolation theory, the
infinite cluster dominates the system and $P_{\infty} \approx p$, i.e. most
of unremoved nodes are connected. Thus, we assume that the small clusters
will have a small effect on $C$ compared to the largest one.  Using this
assumption, Eq.~(\ref{f_def}) can be written as
\begin{equation}
  C\equiv 1-F \equiv \frac{\Sigma^n _{j=1} N_j(N_j-1)}{N(N-1)} \approx
  \frac{N_{\infty}(N_{\infty}-1)}{N(N-1)}\approx \frac{N_{\infty}^2}{N^2}\approx
  P_{\infty}^2.
\label{f_f2}
\end{equation}
Therefore, we expect $P_{\infty}$ and $C$ have the relationship
$P_{\infty}\approx C^{1/2}$ when $p>p_c$ (but not too close to $p_c$). When
$p\le p_c$, the infinite cluster loses its dominance in the system and
$P_{\infty}\sim \mathrm{ln}(N)/N\to 0$ for large $N$~\cite{cohen2}. Here
significant variations between $P_{\infty}$ and $C^{1/2}$ are expected, as
indeed seen in Fig.~\ref{f_vs_f2}.

\section{Simulations}

\begin{figure}[!ht]
\includegraphics[width=0.45\textwidth]{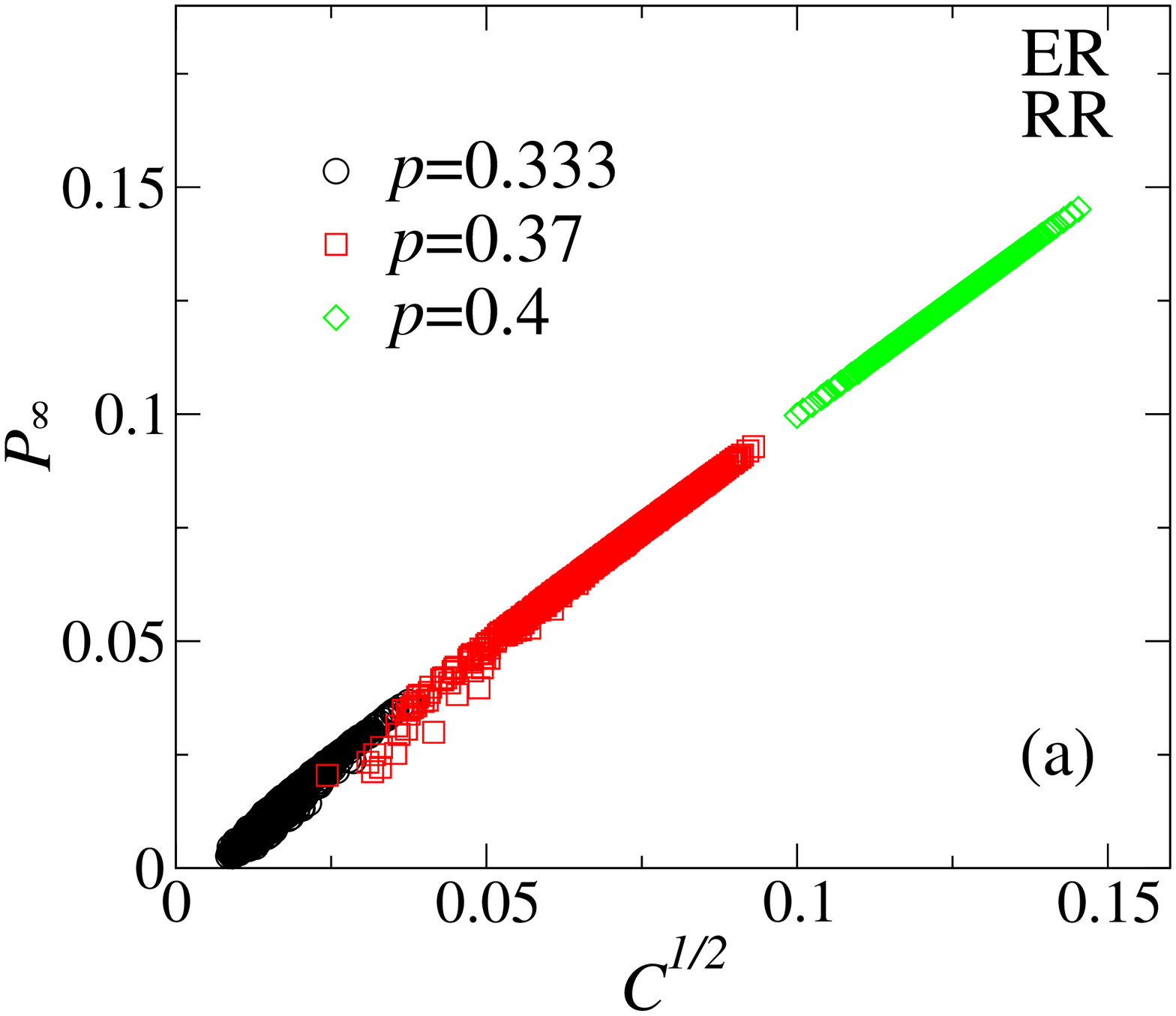}
\includegraphics[width=0.45\textwidth]{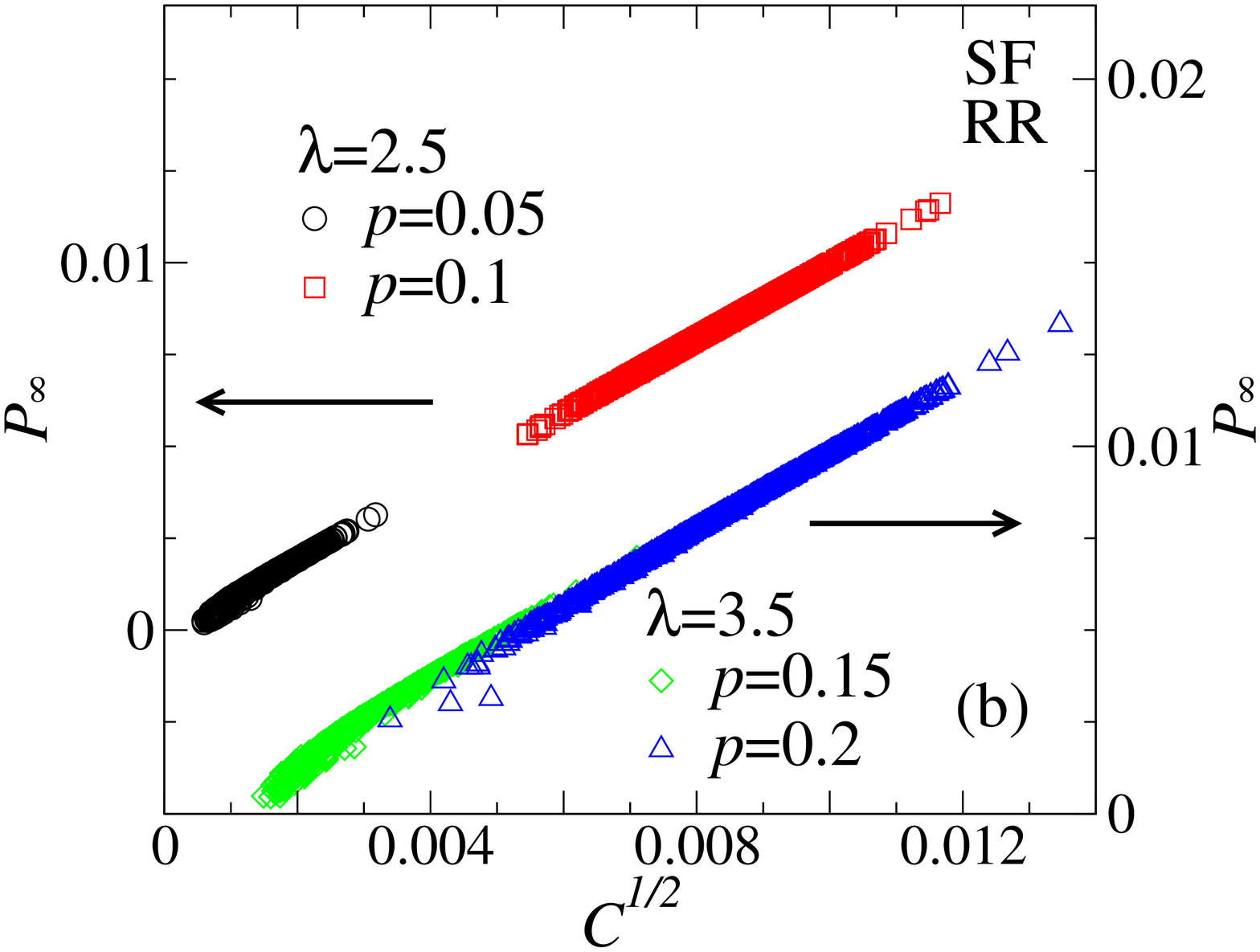}
\includegraphics[width=0.45\textwidth]{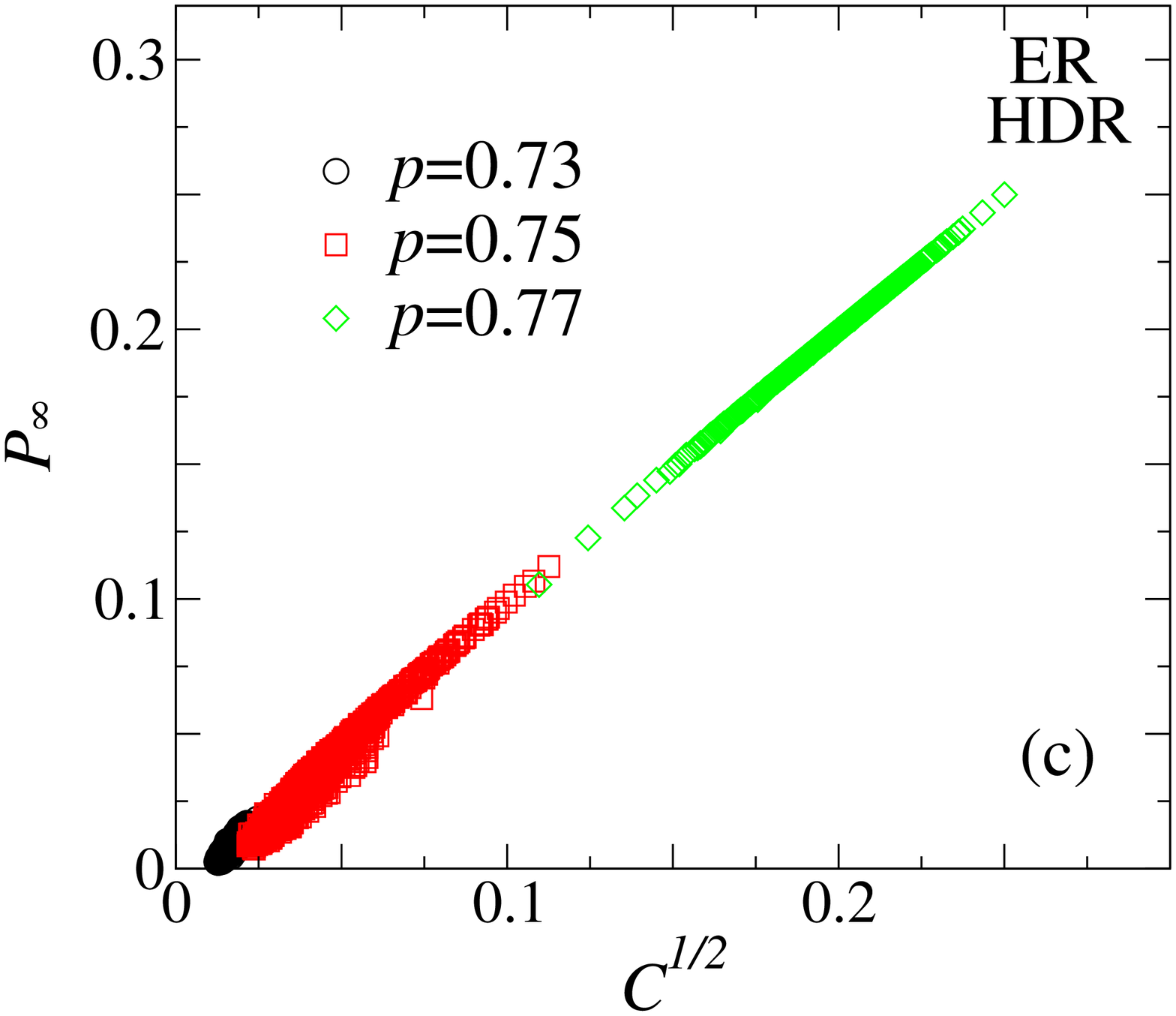}
\includegraphics[width=0.45\textwidth]{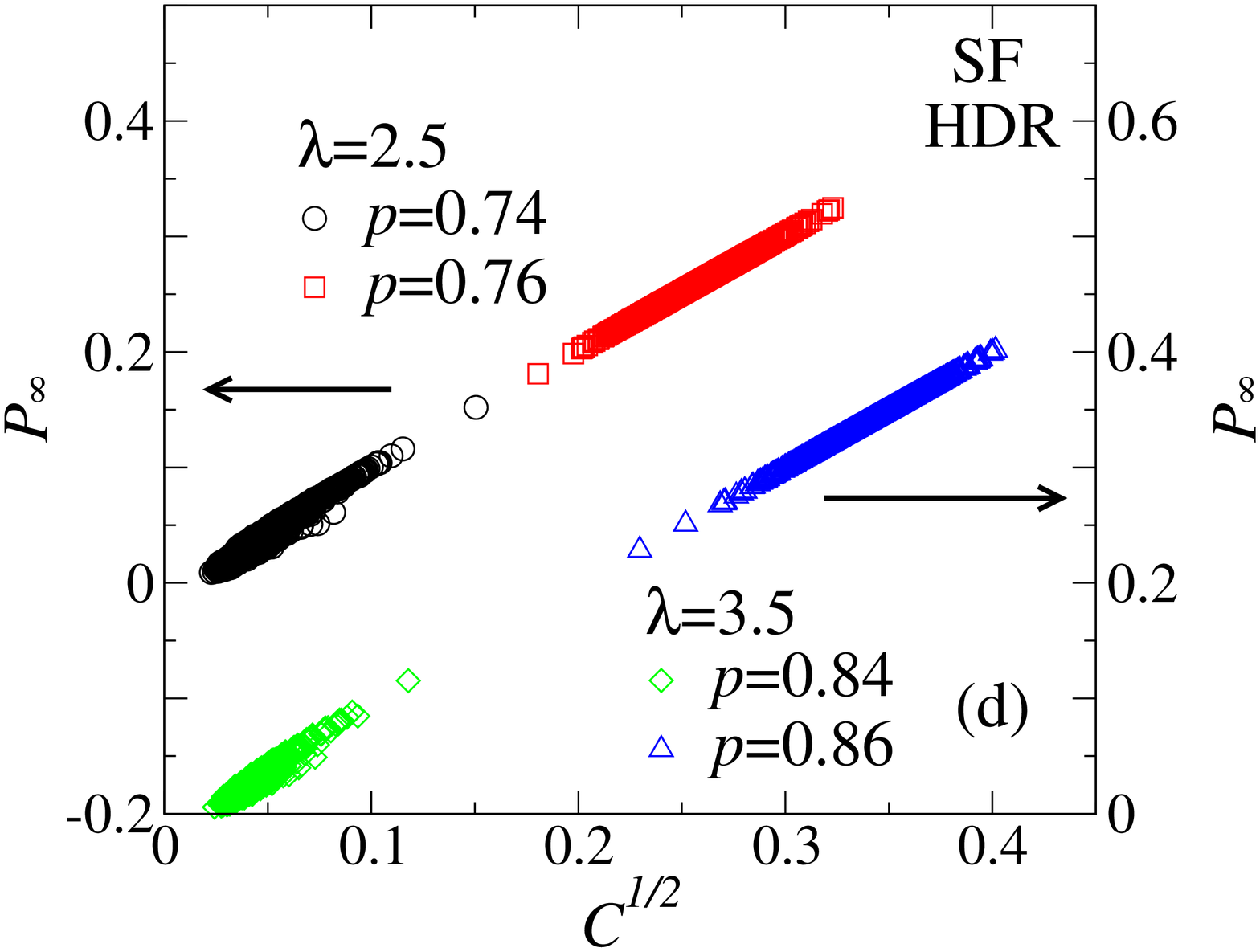}
\includegraphics[width=0.45\textwidth]{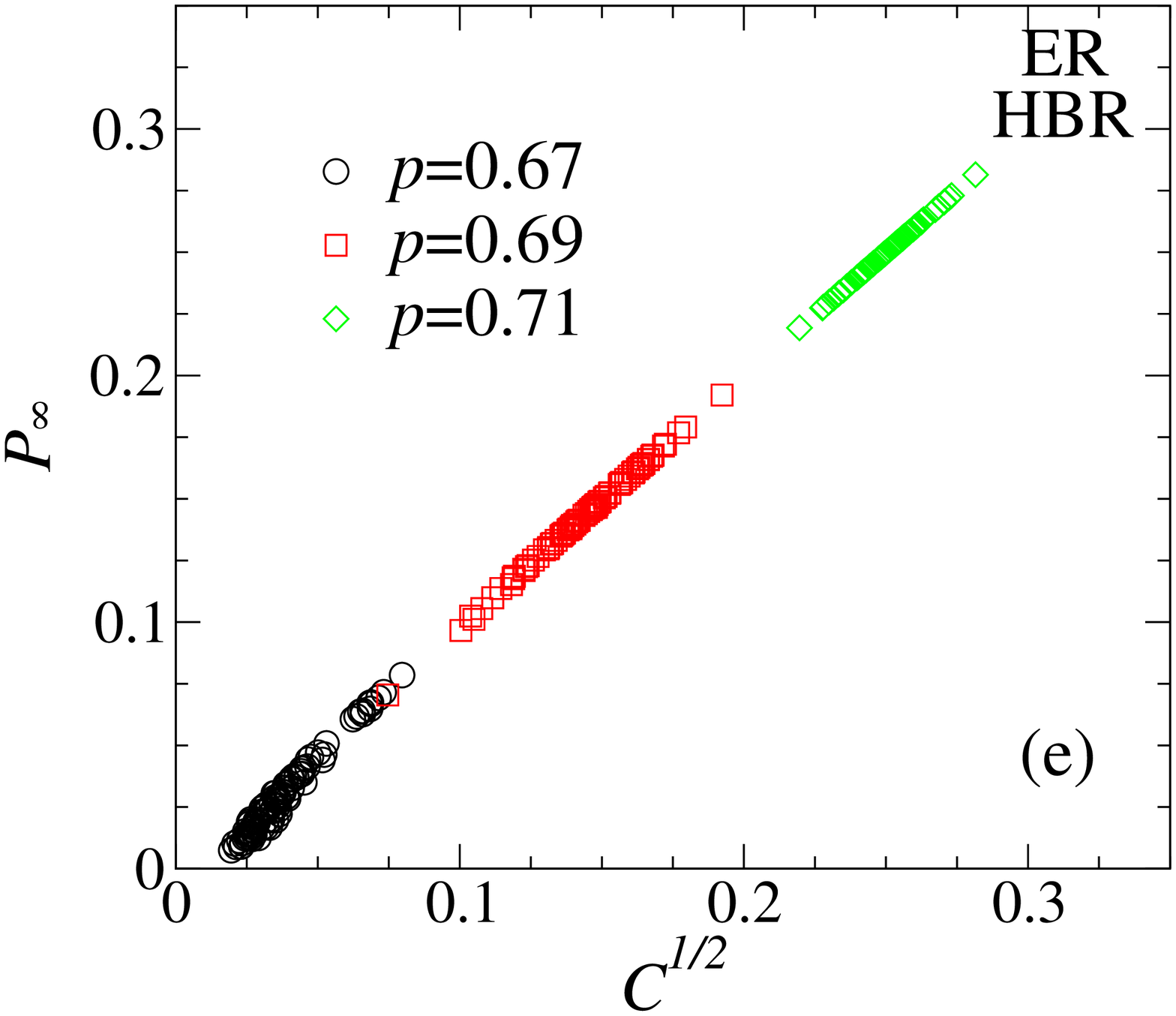}
\includegraphics[width=0.45\textwidth]{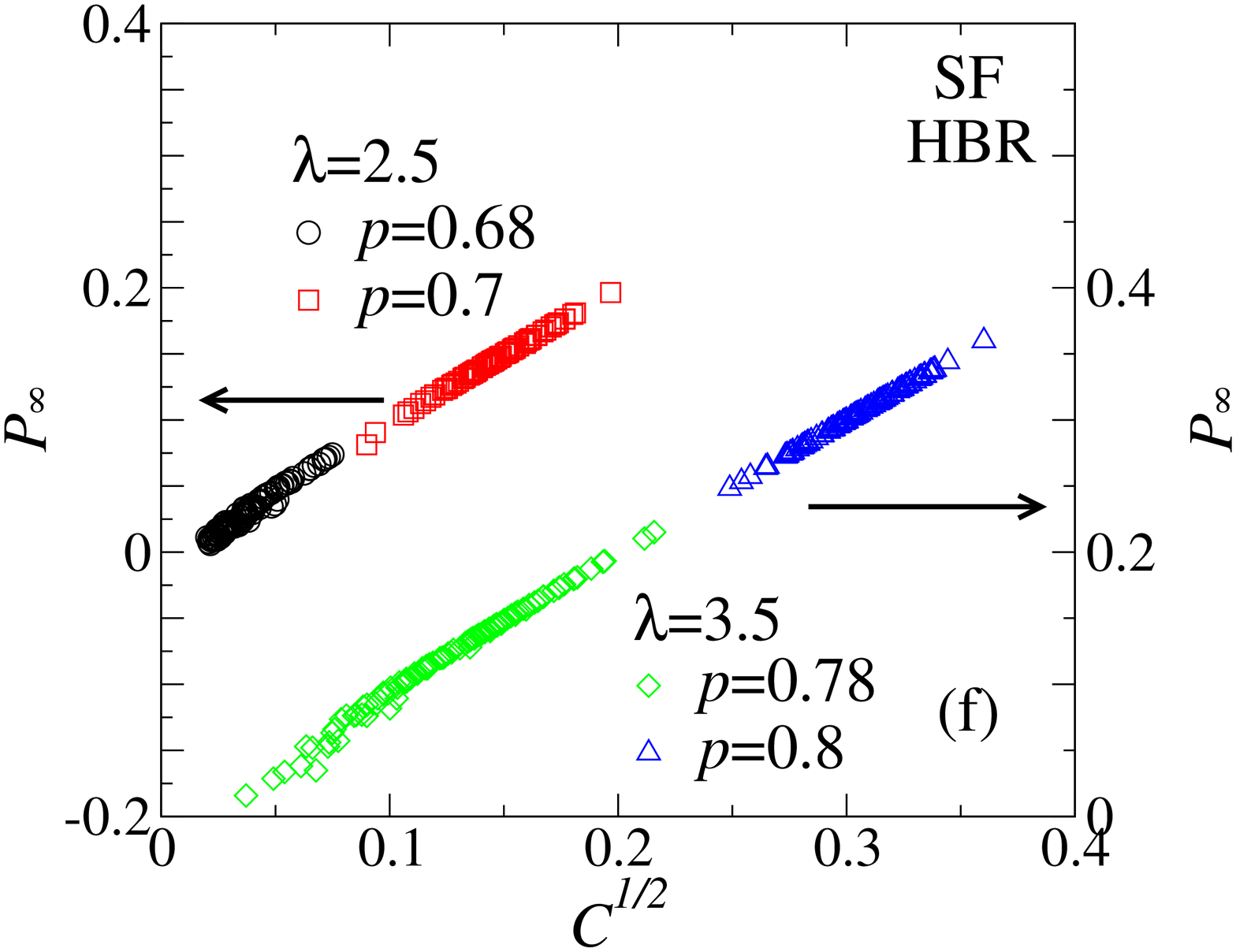}
\caption{Relationship between $C^{1/2}$ and $P_{\infty}$ for ER and SF
  networks with system size $N=50000$. For ER networks, the average degree
  $\langle k \rangle=3$, and for SF networks, $\lambda=2.5$ and $3.5$. The
  graphs are (a) RR strategy on ER networks, (b) RR strategy on SF networks,
  (c) HDR strategy on ER networks, (d) HDR strategy on SF networks, (e) HBR
  strategy on ER networks and (f) HBR strategy on SF networks.
    \label{f_vs_f2}}
\end{figure}

We test by simulations the relationship $C\sim P_{\infty}^2$ derived for
$p>p_c$ in Eq.~(\ref{f_f2}). In Fig.~\ref{f_vs_f2}(a) we plot $P_{\infty}$ vs
$C^{1/2}$ for RR strategy in ER networks and for several values of $p$.  As
predicted by Eq.~(\ref{f_f2}), the plot of $P_{\infty}$ vs $C^{1/2}$ yields a
linear relationship with slope equal to $1$ when $p> p_c=1/\langle
k\rangle=1/3$.  The range of $P_{\infty}$ and $C^{1/2}$ for $p=0.4$ is due to
the variation of $P_{\infty}$ for a given $p$ and the same variation appears
for $C^{1/2}$ showing that the infinite cluster dominates and Eq.~(\ref{f_f2})
is valid.  However, when $p$ drops close to $p_c=1/3$, the system approaches
criticality and the one-to-one correspondence between $C^{1/2}$ and
$P_{\infty}$ is not as strong. This variation is attributed to the presence of
clusters other than the infinite one, which influence $C$ but not
$P_{\infty}$.

Similar behavior is observed for RR strategy in SF networks with
$\lambda=3.5$ shown in Fig.~\ref{f_vs_f2}(b).  For $\lambda=3.5$, the
variation in $C^{1/2}$ emerge close to $p_c=0.2$.  However, for
$\lambda=2.5$, percolation theory suggests that $p_c$ approaches $0$ for
large systems. As a result, no significant variation is observed even when
$P_{\infty}$ is as small as $5\cdot 10^{-4}$.  This observation supports that
the SF networks with $\lambda<3$ are quite robust in sustaining its infinite
cluster against random removal~\cite{cohen}.  Figs.~\ref{f_vs_f2}(c),
\ref{f_vs_f2}(d),\ref{f_vs_f2}(e) and \ref{f_vs_f2}(f) show the results for
HDR and HBR strategies in ER and SF networks.  For these targeted strategies,
the variation of $C^{1/2}$ and $P_{\infty}$ shows up at significantly higher
$p$ compared to the random case, indicating that the infinite cluster breaks
down easier under HDR and HBR attacks for both ER and SF networks, as seen
also in Fig.~\ref{f_kill}. At this point, the SF network with $\lambda=2.5$
becomes no longer as robust as in the random case, as can be clearly observed
in the large variation at $P_{\infty}\approx 0.05$.

\begin{figure}[!ht]
  \includegraphics[width=0.45\textwidth]{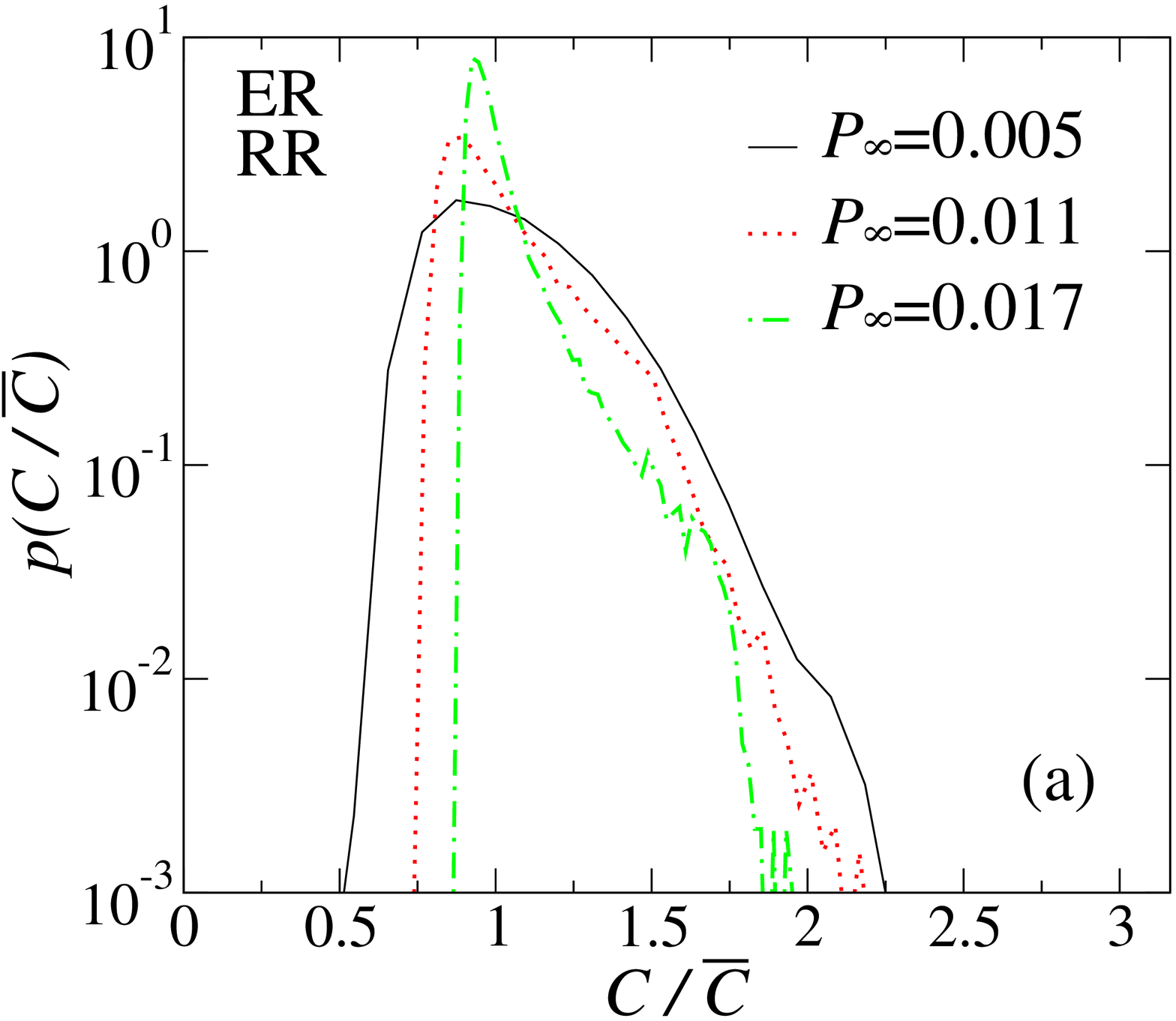}
  \includegraphics[width=0.45\textwidth]{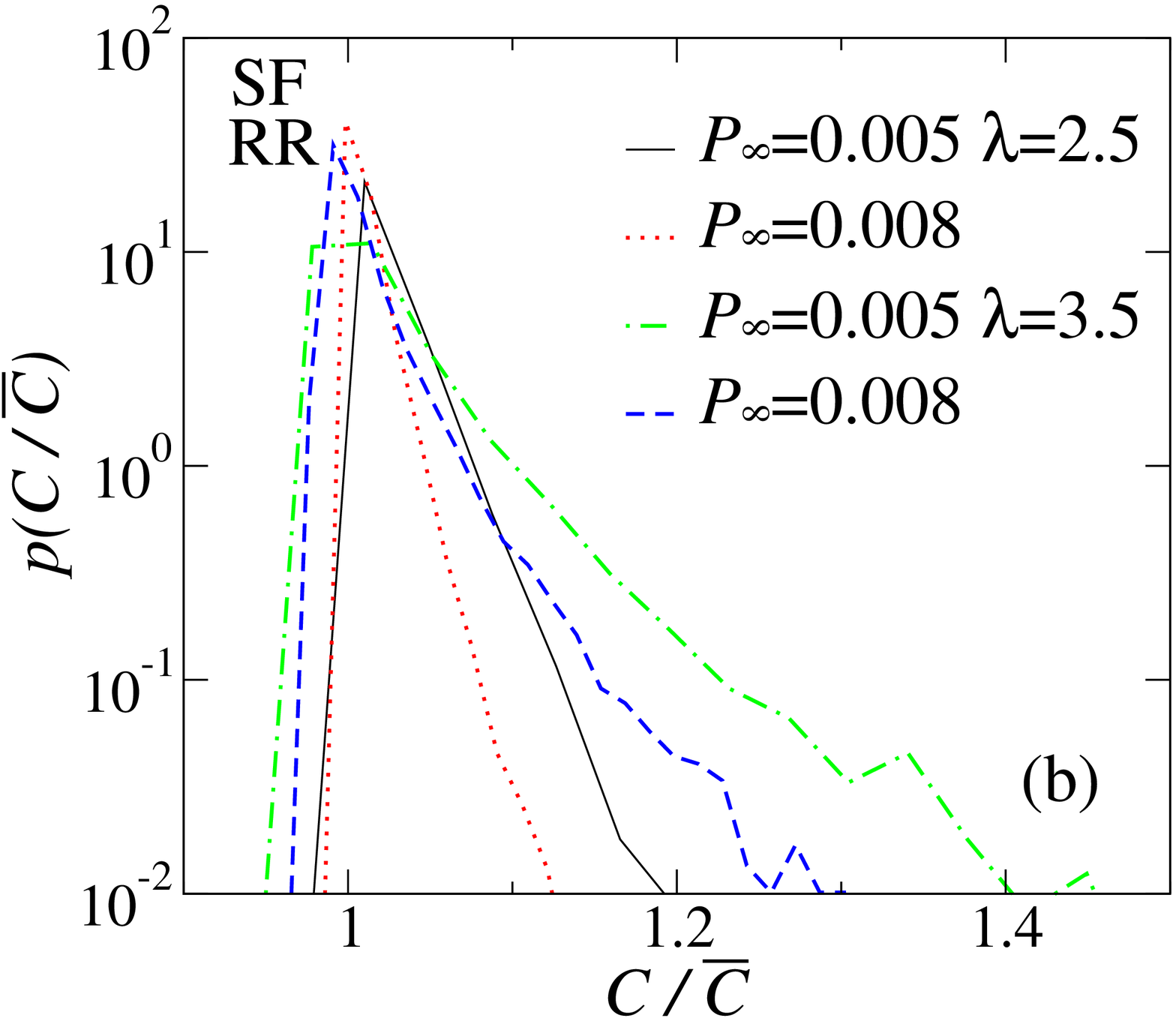}
  \includegraphics[width=0.45\textwidth]{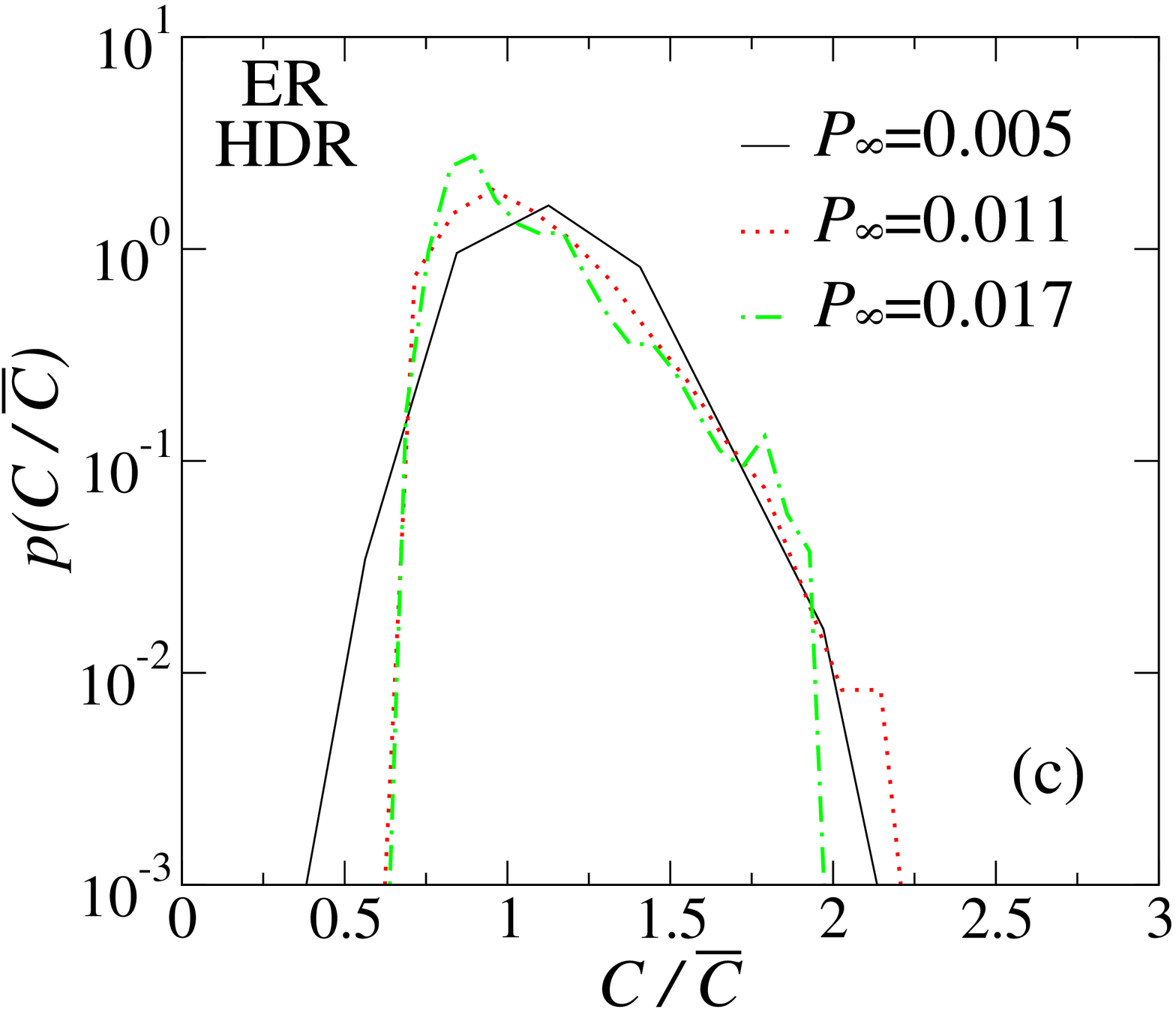}
  \includegraphics[width=0.45\textwidth]{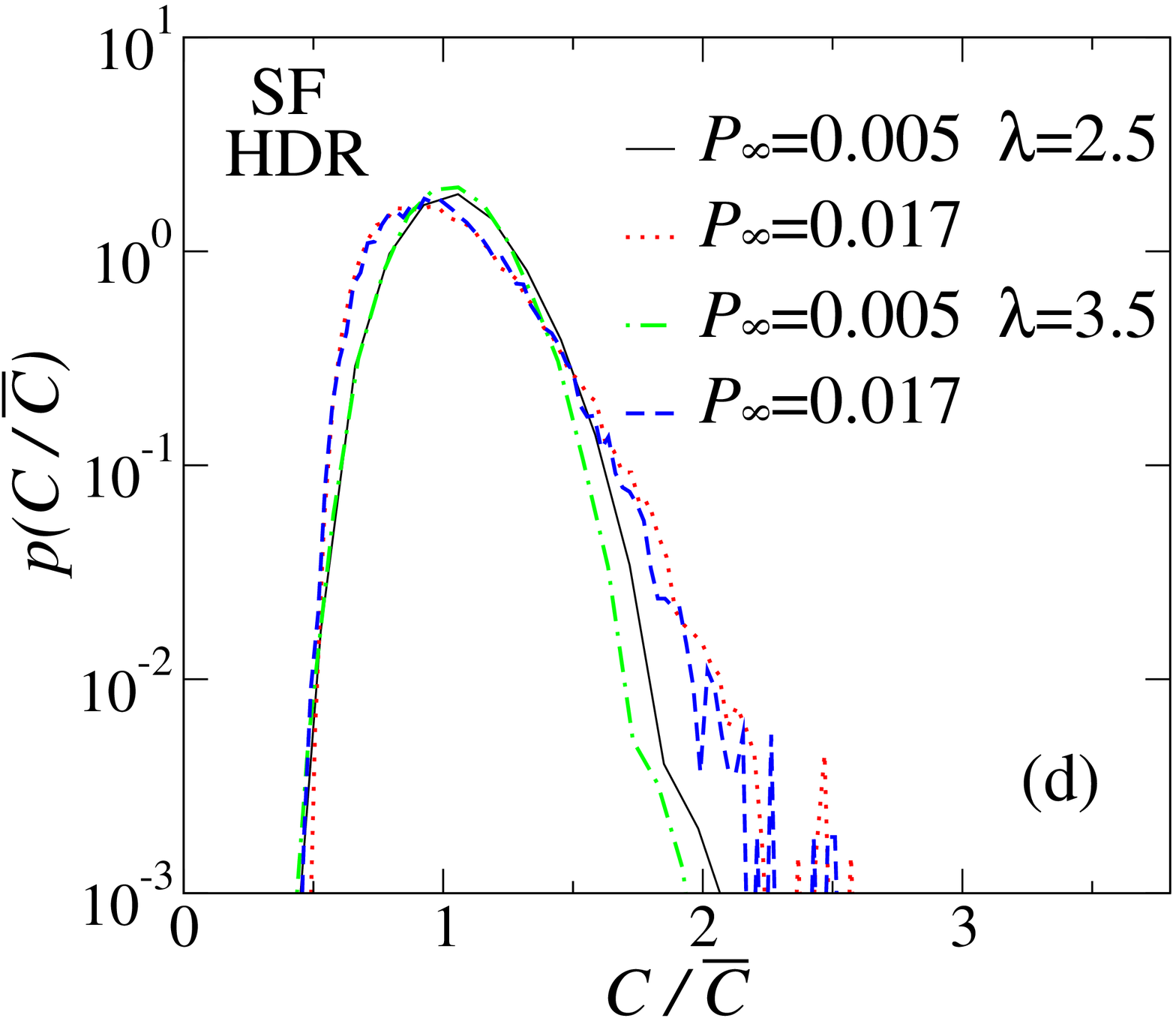}
  \caption{Probability distributions $p(C/\bar{C})$ versus $C/\bar{C}$ for
    several values of $P_{\infty}$ and for ER networks with $\langle k
    \rangle =3$, $N=200000$ and SF networks with $N=80000$ and $\lambda=2.5$
    and $3.5$. (a) RR strategy on ER networks, (b) RR strategy on SF
    networks, (c) HDR strategy on ER networks and (d) HDR strategy on SF
    networks.\label{f_limit}}
\end{figure}

To further investigate the characteristics of the variation of $C$ for a
given $P_{\infty}$, we calculate the probability distributions $p(C)$ versus
$C/\bar{C}$ for a given $P_{\infty}$ where $\bar{C}$ is the average value of
$C$ and the results are plotted in Fig.~\ref{f_limit}. In this case, $C^*$,
the most probable value of $C$, is determined by the fixed infinite cluster
size $P_{\infty}$ with $C^*\approx P_{\infty}^2$, and the broadness of $p(C)$
comes from presence of clusters other than the infinite one.  Because the
largest cluster size is fixed, the upper cutoff of $p(C)$ emerges due to the
limitation on the sizes of other clusters that by definition must be smaller
than the largest cluster.  For the RR strategy, the broadness of $p(C)$ for
ER network is bigger than that of SF networks at the same $P_{\infty}$,
especially for $\lambda =2.5$ where the system is always high above
criticality and the variation is relatively small.  On the contrary, for the
HDR and HBR strategies, the broadness of $p(C)$ for ER and SF networks are of
the same order due to the fact that for HDR and HBR, $p_c$ is also finite for
$\lambda=2.5$.  This observation is consistent with the results shown in
Fig.~\ref{f_vs_f2}.

\begin{figure}[!ht]
\includegraphics[width=0.45\textwidth]{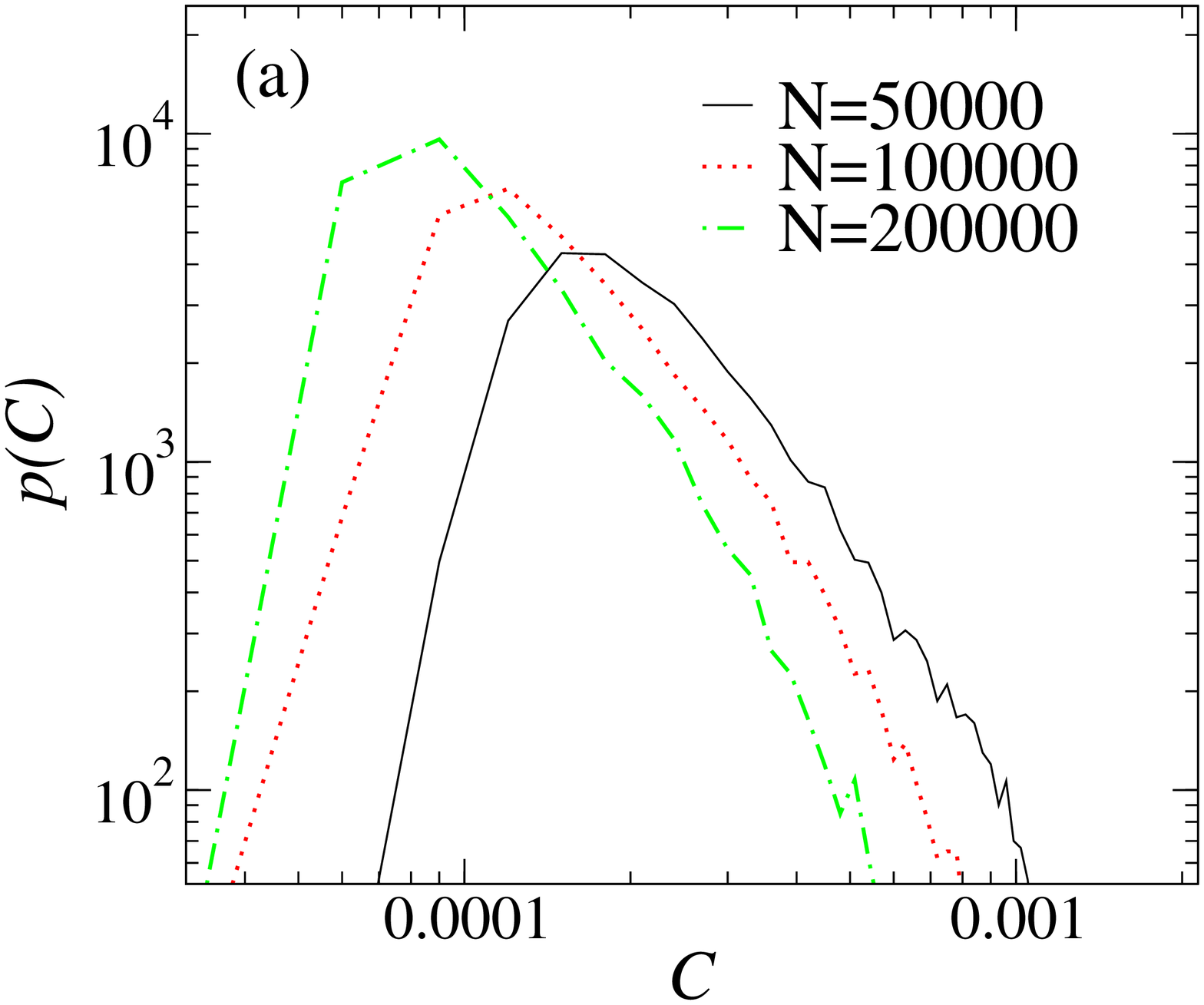}
\includegraphics[width=0.45\textwidth]{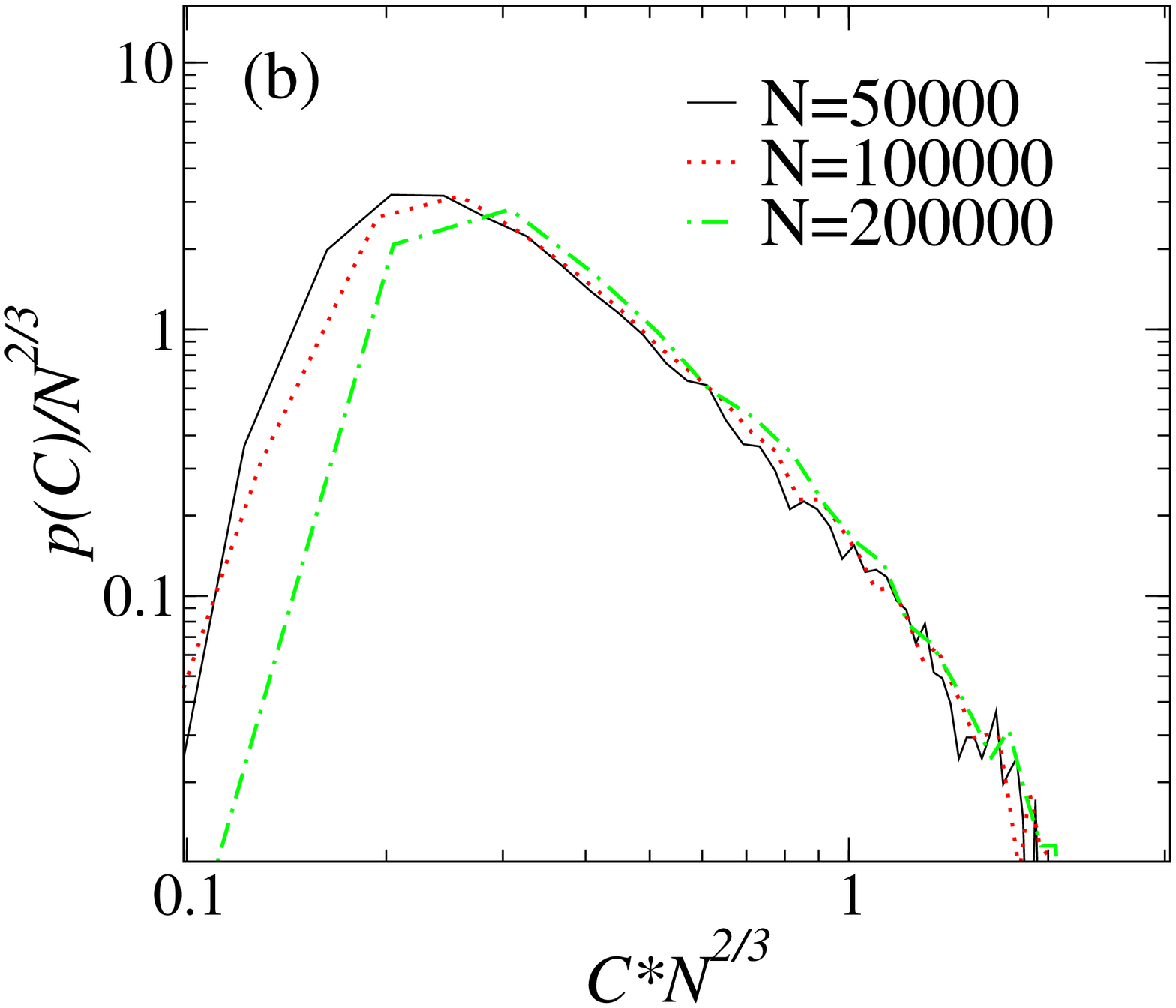}
\caption{The dependence of $p(C)$ on the system size $N$ with $p=p_c$ for (a) before scaling and (b) after scaling.
  Simulations are performed on ER networks with $\langle k \rangle=3$.
  \label{f_n}}
\end{figure}

Now we focus on the dependence of $p(C)$ on the system size $N$ at $p_c$
(Fig.~\ref{f_n}). From percolation theory and for ER under RR strategy, the
infinite cluster size $N_{\infty}$ at criticality behaves
as~\cite{erdos,cohen4}
\begin{equation}
N_{\infty}\sim N^{2/3}.
\end{equation}
Since $C$ follows similar behavior as $N_{\infty}$ at criticality, we
expect $C$ for $p=p_c$ to behave as,
\begin{equation}
C\equiv 1-F \approx (N_{\infty}/N)^2\sim N^{-2/3}.
\label{eq_f_n}
\end{equation}
Thus, we expect the probability distribution $p(C)$ with $p=p_c$ to scale as
\begin{equation}
 p(C)=N^{2/3}g(C N^{2/3})
\label{eq_pf_n}
\end{equation}
where $g$ is a scaling function.

Fig.~\ref{f_n}b supports this scaling relationship. We calculate $p(C)$ for
RR strategy at criticality on ER networks with $N$ values of $50000$,
$100000$, $200000$ and $\langle k \rangle=3$ (shown in Fig.~\ref{f_n}a), and
~\cite{shai}find a good collapse when plotted (Fig.~\ref{f_n}b) using the scaling form of
Eq.~(\ref{eq_pf_n}).

\section{Real networks}

\begin{figure}[!ht]
\includegraphics[width=0.45\textwidth]{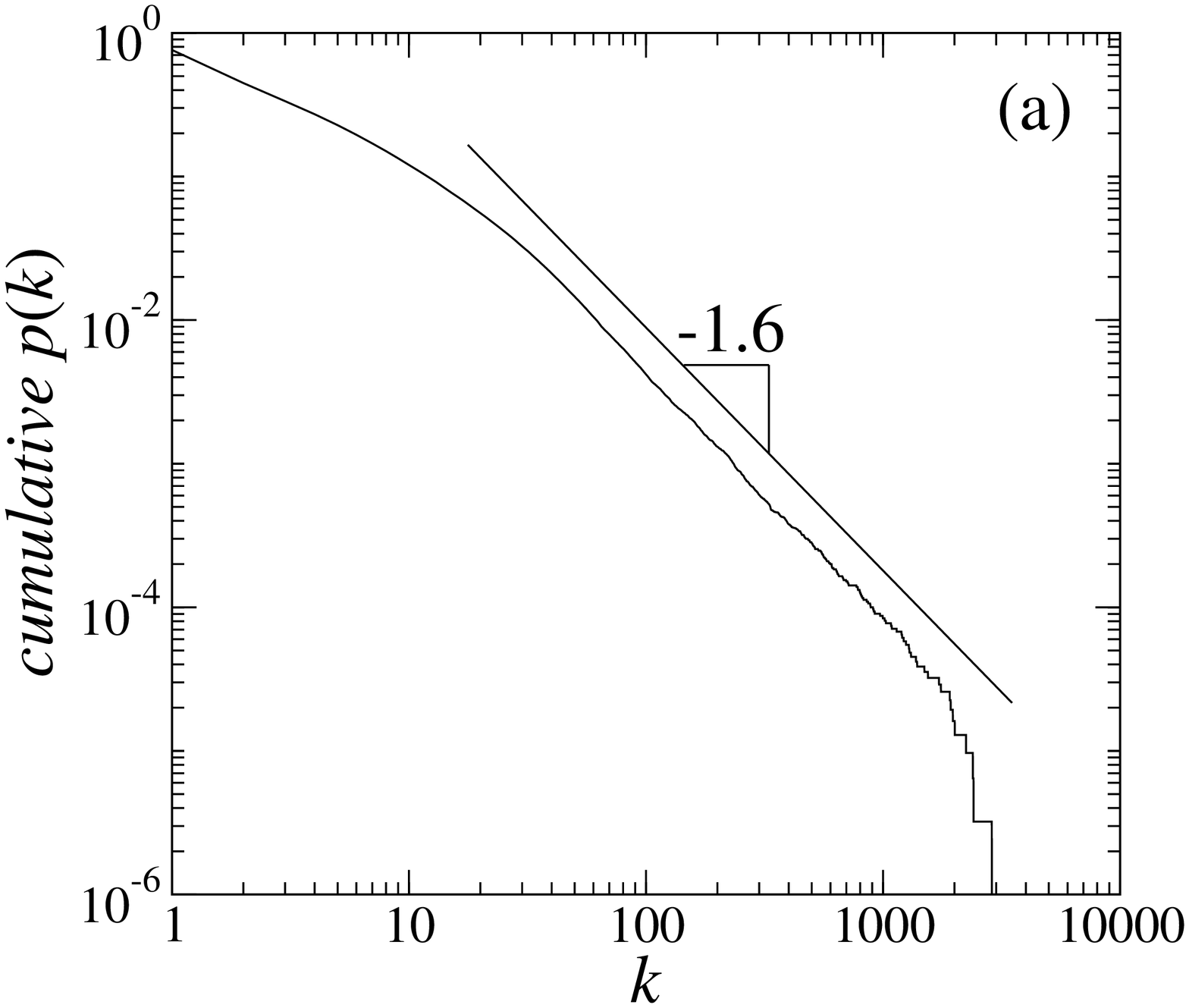}
\includegraphics[width=0.45\textwidth]{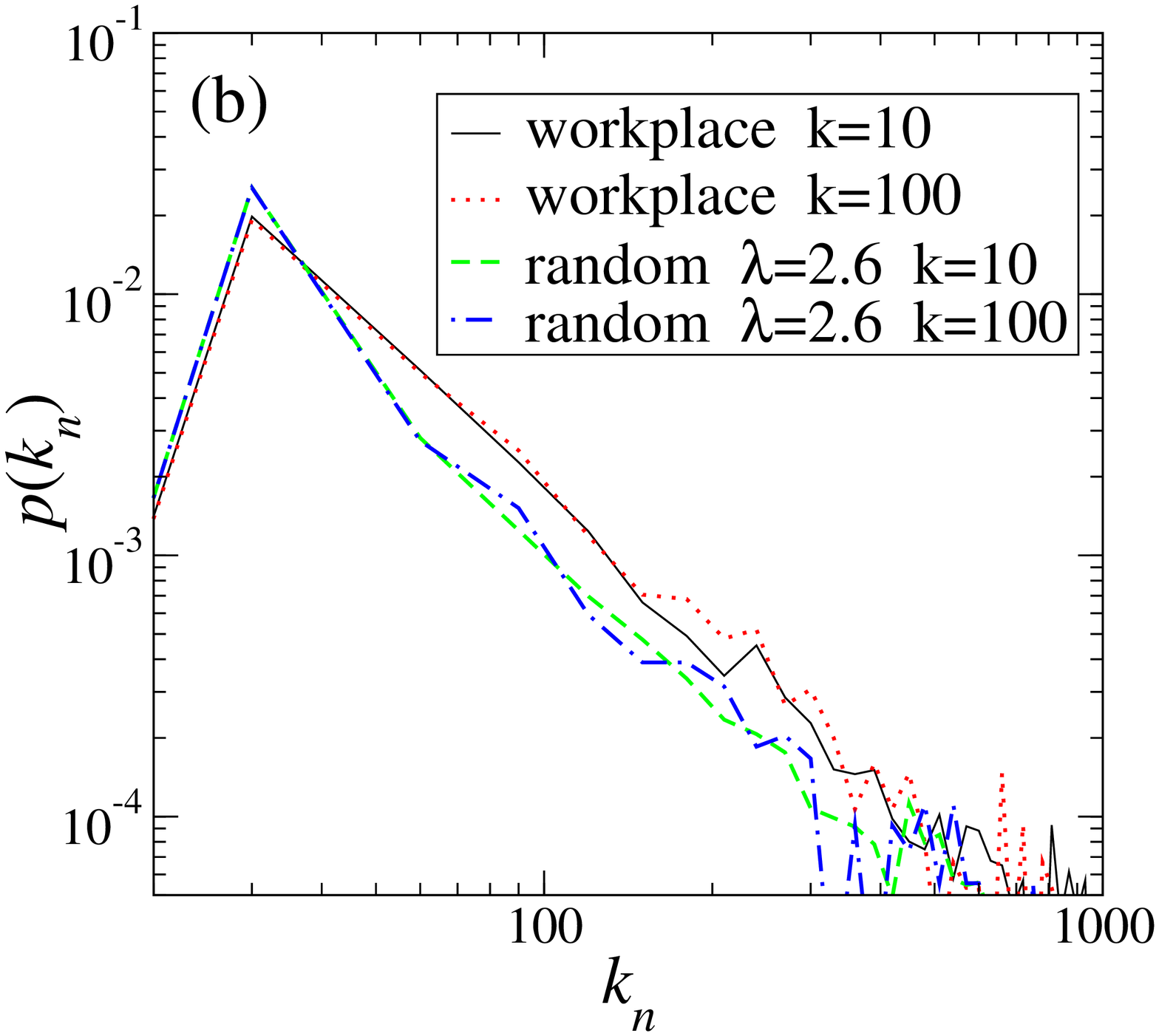}
\includegraphics[width=0.45\textwidth]{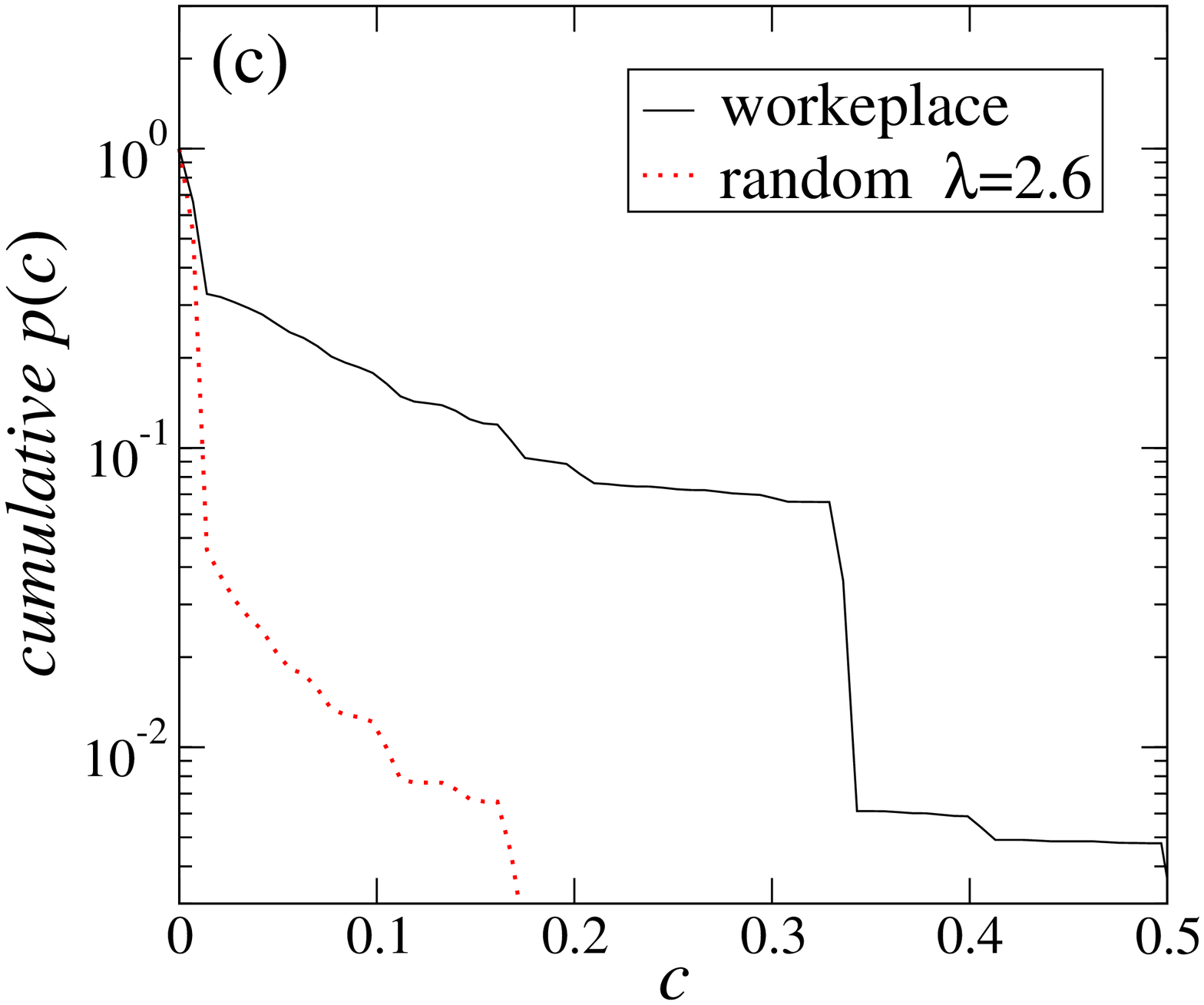}
\includegraphics[width=0.45\textwidth]{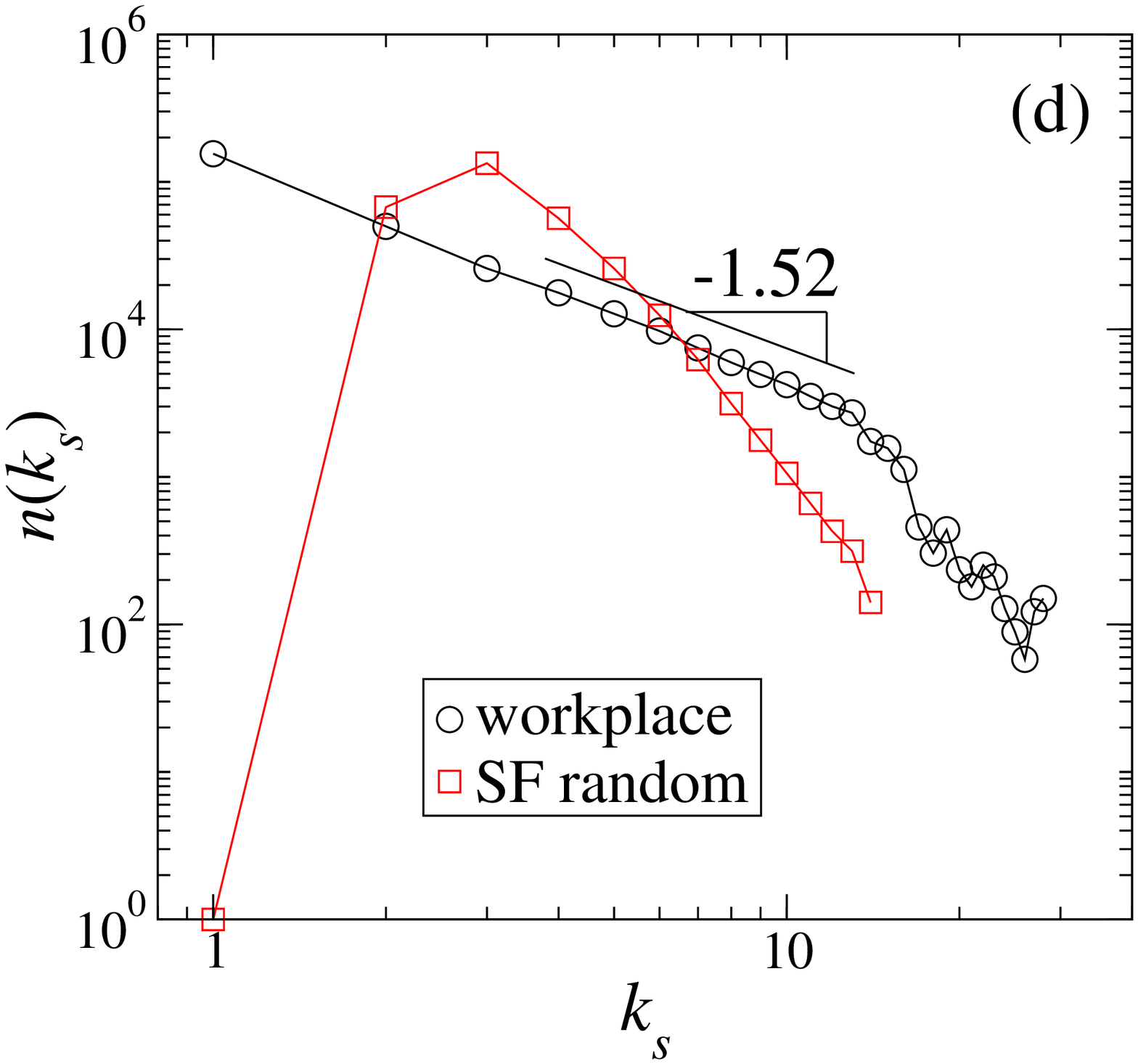}
\includegraphics[width=0.45\textwidth]{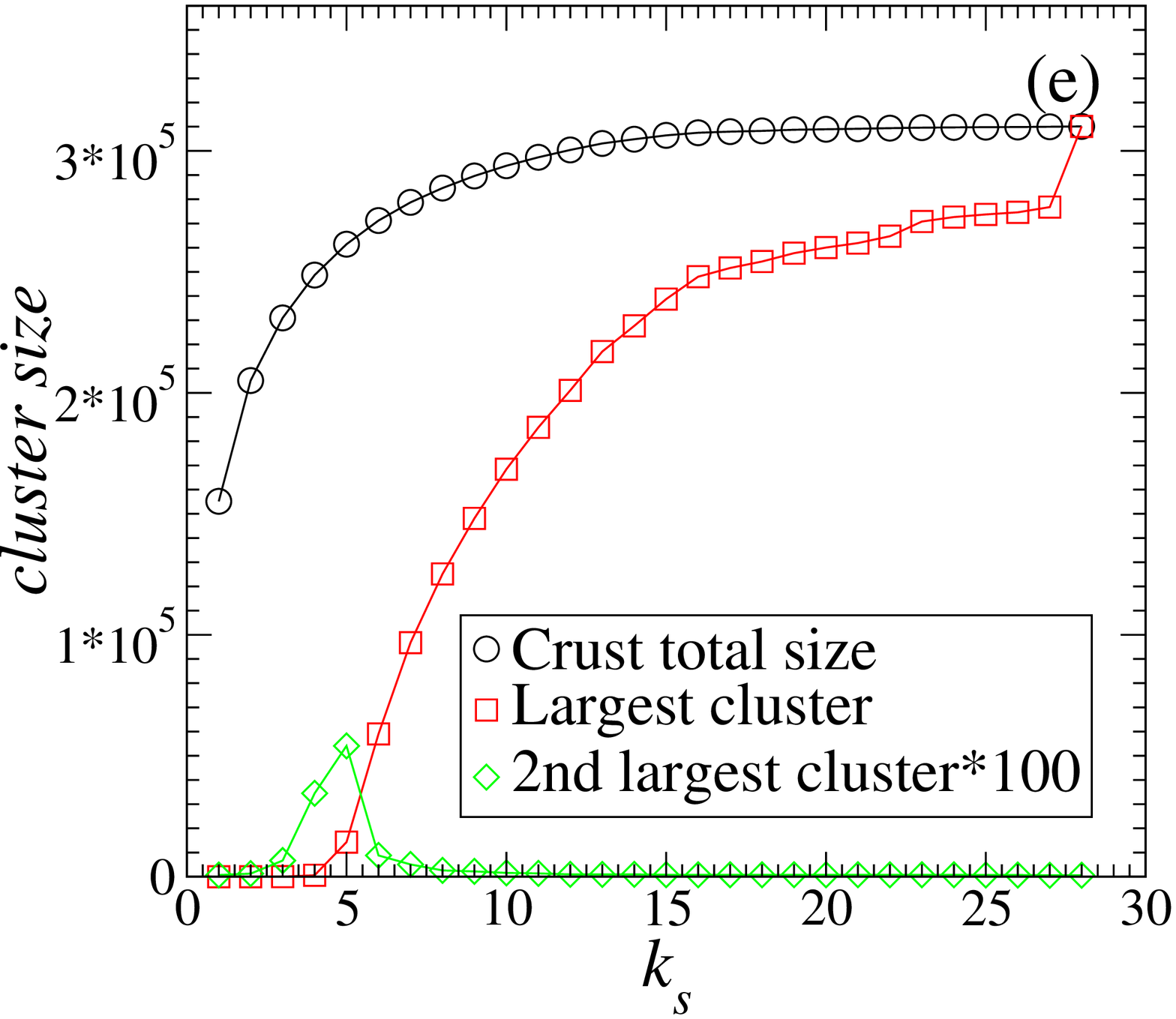}
\caption{Properties of the Swedish network of workplaces. (a) The cumulative
  degree distribution (showing $\lambda=2.6$). (b) The distribution of $k_n$,
  the degree of the neighbors of nodes having degree $k$. (c) The cumulative
  distribution of clustering coefficient $c$. (d) Number of nodes in shell
  $k_s$. (e) Size of largest and second largest cluster in each k-crust. In
  (b), (c) and (d) the distributions of random SF networks with the same
  $\lambda$ and $N$ are plotted for comparison.\label{real_properties}}
\end{figure}

\begin{figure}[!ht]
\includegraphics[width=0.45\textwidth]{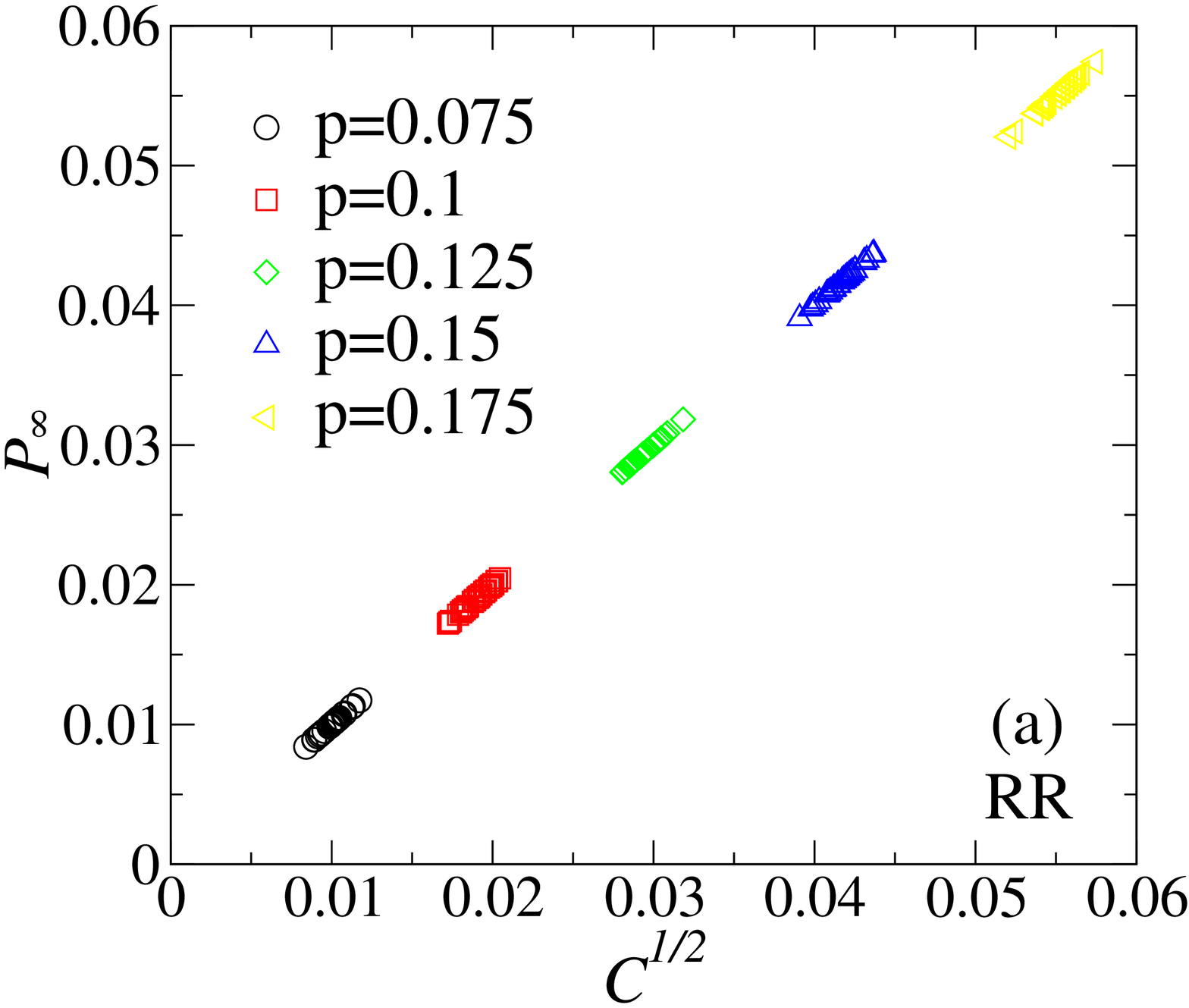}
\includegraphics[width=0.45\textwidth]{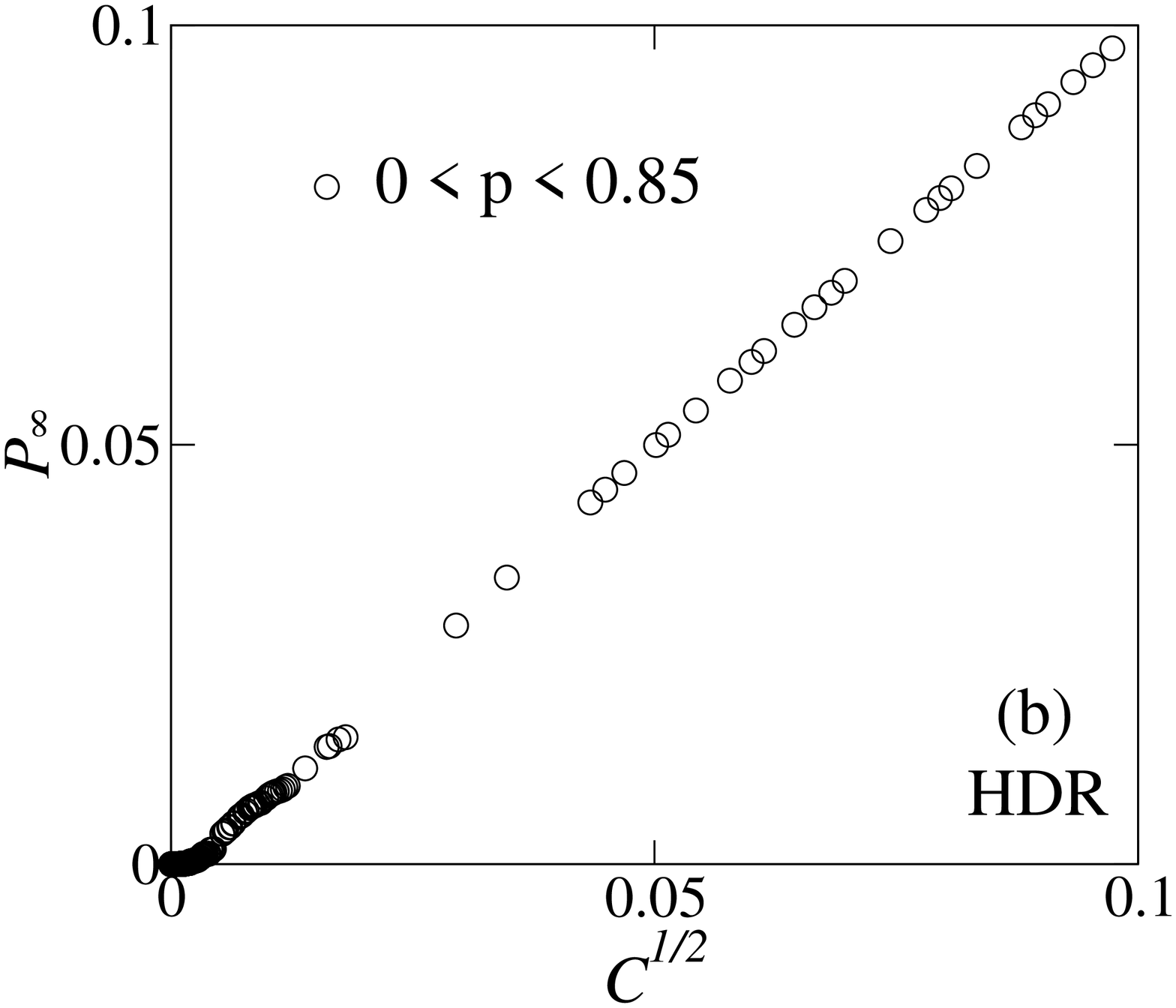}
\includegraphics[width=0.45\textwidth]{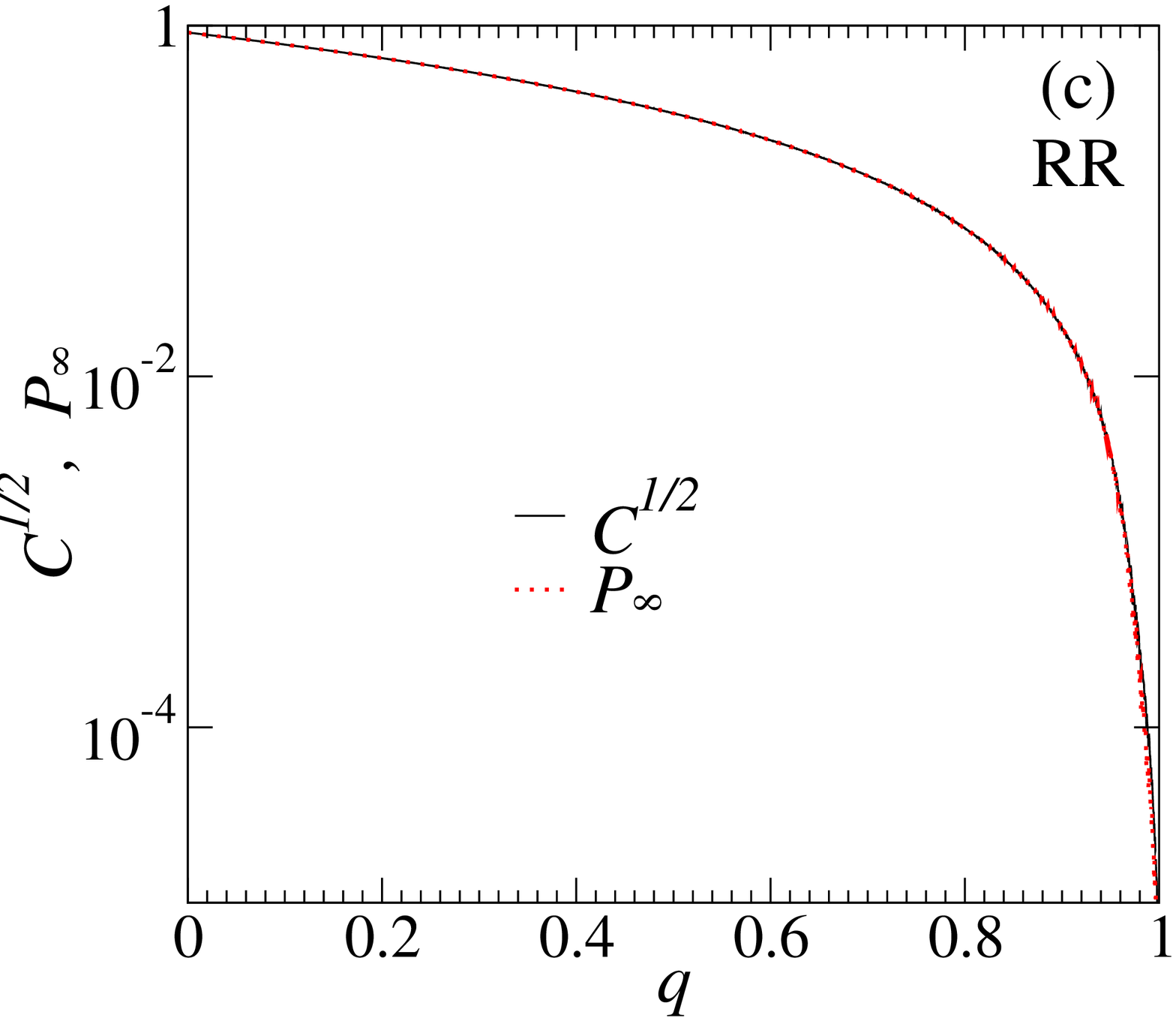}
\includegraphics[width=0.45\textwidth]{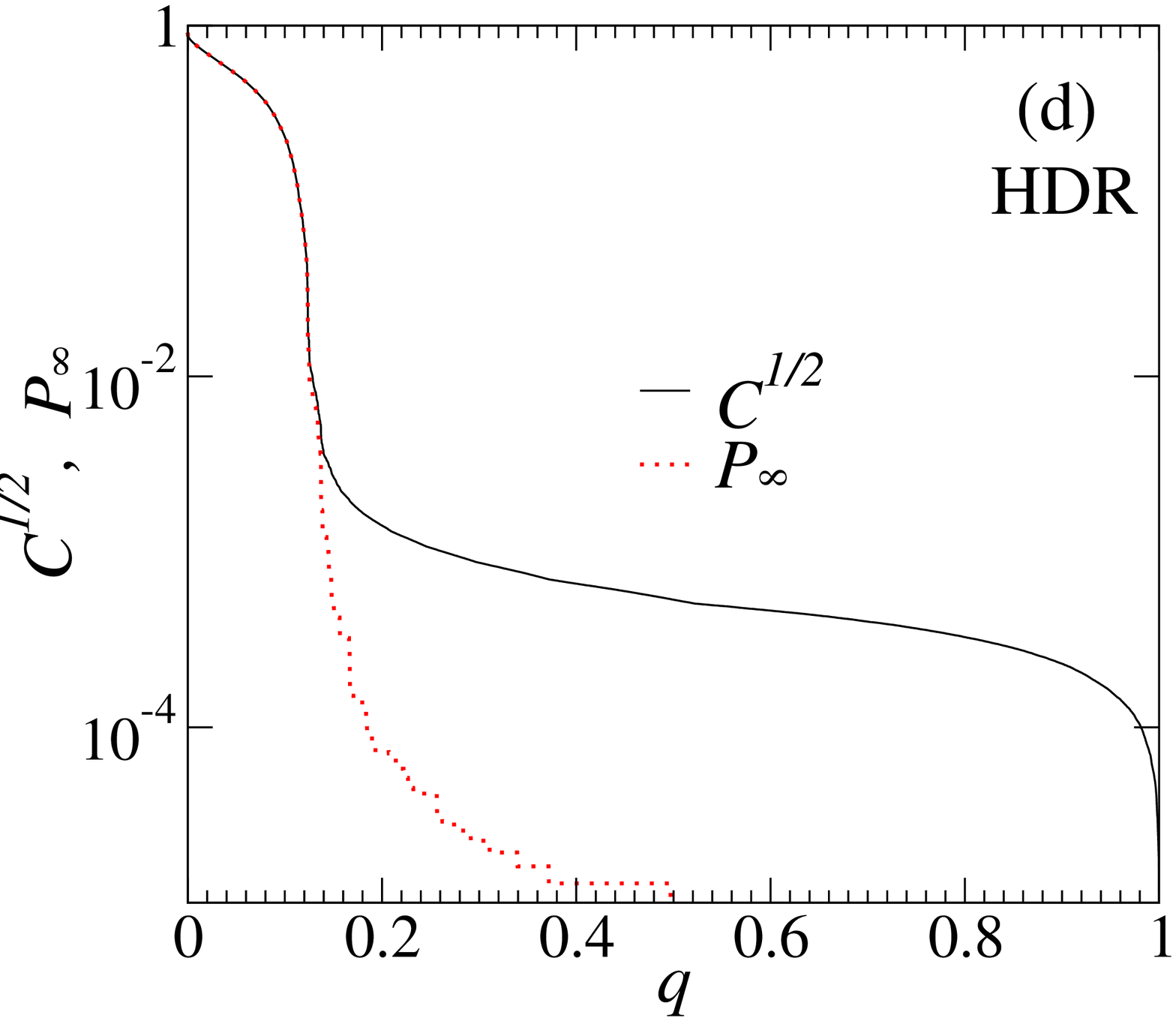}
\caption{$P_{\infty}$ vs $C^{1/2}$ for (a) RR strategy and (b) HDR strategy
  and plot $C^{1/2}$, $P_{\infty}$ versus $q$ for (c) RR strategy and (d) HDR
  strategy for the Swedish network of workplaces with $N=310136$
  nodes.\label{f_real}}
\end{figure}

The ER networks and the SF networks that we have been studying are random
ensemble of networks which are only determined by their degree distribution.
It is known that many real networks often exhibit important structural
properties relevant for percolation properties such as high level of
clustering, assortativity and fractality that random networks do not
exhibit~\cite{song, newman2}.  We therefore test our results about the
relation between $C$ and $P_{\infty}$ on an example of a large real social
network. The network we use is extracted from a data set obtained from
Statistics Sweden~\cite{scb} and consists of all geographical workplaces in
Sweden that can be linked with each other by having at least one employee
from each workplace sharing the same household.  Household is defined as a
married couple or a couple having kids together that are living in the same
flat or house. Unmarried couples without kids and other individuals sharing
household are not registered in the dataset as households.  This kind of
network have been shown to be of importance for the spreading of
Influenza~\cite{viboud} and are also likely to be important for spreading of
information and rumors in society. The network consists of $310136$ nodes
(workplaces) and $906260$ links (employees sharing the same households) and,
as shown in Fig.~\ref{real_properties}(a), is approximately a SF network with
$\lambda\approx 2.6$ and an exponential cut off. The network shows almost no
degree-degree correlation (assortativity)~(Fig.~\ref{real_properties}(b)).
However, the workplace network clustering coefficient $c$ is significantly
higher than that of a random SF network with same $\lambda$ and
$N$~(Fig.~\ref{real_properties}(c)). The average of $c$ is $0.048$ for the
workplace network versus $3.2\times 10^{-4}$ for the random SF networks,
which is consistent with the earlier social network
studies~\cite{gabor,konstantin}.  Fig.~\ref{real_properties}(d) shows the
node distribution $n(k_s)$ of k-shell ($k_s$) in the network compared to that
of a random SF network with same $\lambda$ and $\langle
k\rangle$~\cite{shai}. It is seen that in the workplace network there exist
significantly more shells and the large shells are more occupied compared to
random SF. The distribution $n(k_s)$ shows a power-law behavior with slope
$-1.52$. This indicates the structure of this real network.
Fig.~\ref{real_properties}(e) shows the crust total size, the largest cluster
size and the second largest cluster size as a function of shell $k_s$. It is
seen that the largest cluster has two transitions. One around $k_s=5$ and the
other at $k_s=27$. At $k_s>5$, the largest cluster increase from zero to a
finite fraction of the network.  This transition is related to the HDR seen
in Fig.~\ref{f_real}(d) (see also~\cite{cohen2}). The second transition at
$k_s=27$ defines the nucleus of the workplace network which include about 100
nodes (see Fig.~\ref{real_properties}(d), $n(28)\approx 100$) which are well
connected to each other. The jump of the largest cluster from $k_s=27$ to
$k_s=28$ from $2.8 \times 10^5$ nodes to $3.1\times 10^5$ nodes (i.e.
$3\times 10^4$ nodes) is due to nodes which are connected only to the
nucleus.  These nodes are called dendrites.  Fig.~\ref{real_properties}(e) is
very similar to the Medusa model~\cite{shai} suggested for the AS
topology of the Internet. Figs.~\ref{f_real}(a) and \ref{f_real}(b) show
simulation results for several values of $p$ for $P_{\infty}$ vs $C^{1/2}$.
The curves are linear, similar to Fig.~\ref{f_vs_f2} for our model networks.
Moreover, Figs.~\ref{f_real}(c) and (d) show that $C^{1/2}$ and $P_{\infty}$
are almost identical above the criticality threshold $p_c$ for a typical
configuration after both RR and HDR.  For $p$ below criticality, differences
appear which are especially obvious for HDR strategy where $q_c = 1-p_c$ is
relatively small.  While $P_{\infty}$ rapidly decreases to a very small value
(below $10^{-5}$), a plateau shows up in the curve of $C^{1/2}$ due to the
influence of the small clusters.

\section{Summary}

In summary, we study the measure for fragmentation $F\equiv 1-C$ proposed in
social sciences and relate it to the traditional $P_{\infty}$ used in physics
in percolation theory. For $p$ above criticality, $C$ and $P_{\infty}$ are
highly correlated and $C\approx P_{\infty}^2$. Close to criticality, for
$p\ge p_c$ and below $p_c$, variations between $C$ and $P_{\infty}$ emerge
due to the presence of the small clusters. For systems close or below
criticality, $F$ gives better measure for fragmentation of the whole system
compared to $P_{\infty}$. We study the probability distribution $p(C)$ for a
given $P_{\infty}$ and find that $p(C)$ at $p=p_c$ obeys the scaling
relationship $p(C)=N^{2/3}g(C N^{2/3})$ for both RR strategy on ER network,
and for HDR on scale free networks.

We thank ONR, European NEST project DYSONET, and Israel Science Foundation
for financial support.

The study was approved by the Regional Ethical Review board in Stockholm
(record 2004/2:9).

\begin{thebibliography}{99}
\bibitem{albert} R. Albert, H. Jeong, and A.-L. Barab\'{a}si, Nature London
  \textbf{406}, 378 (2000).
\bibitem{pastor2}R. Pastor-Satorras and A. Vespignani, {\it Evolution and
    Structure of the Internet: A Statistical Physics Approach} (Cambridge
  University Press, Cambridge, England, 2004).
\bibitem{doro} S.N. Dorogovtsev and J.F.F. Mendes, {\it Evolution of
    Networks: From Biological Nets to the Internet and WWW} (Oxford
  University Press, Oxford, 2003).
\bibitem{bocca} S. Boccaletti, V. Latora, Y. Moreno, M. Chavez, D.-U. Hwang,
  Physics Reports \textbf{424}, 175 (2006).
\bibitem{paxon} V. Paxon, IEEE/ACM Trans. Networking \textbf{5}, 601 (1997).
\bibitem{cohen} R. Cohen, K. Erez, D.ben-Avraham, and S. Havlin, Phys. Rev.
  Lett. \textbf{85}, 4626 (2000).
\bibitem{callaway} D.S. Callaway, M.E.J. Newman, S.H. Strogatz, and
  D.J. Watts, Phys. Rev. Lett. \textbf{85}, 5468 (2000).
\bibitem{cohen2} R. Cohen, et al., Phys. Rev. Lett. \textbf{86}, 3682 (2001).
\bibitem{valente} A.X.C.N. Valente, A. Sarkar, and H.A. Stone,
  Phys. Rev. Lett. \textbf{92}, 118702 (2004).
\bibitem{gerry} G. Paul, T. Tanizawa, S. Havlin, and H.E. Stanley,
  Eur. Phys. J. B \textbf{38}, 187 (2004).
\bibitem{chung} F. Chung and L. Lu, Ann. Combinatorics \textbf{6}, 125 (2002).
\bibitem{burda} Z. Burda and A. Krzywicki, Phys. Rev. E \textbf{67}, 046118
  (2003).
\bibitem{song} C. Song et al. Nature \textbf{433}, 392 (2005).
\bibitem{freeman} L.C. Freeman, {\it The Development of Social Network
    Analysis: A Study in the Sociology of Science} (Empirical, 2004).
\bibitem{wasserman} S. Wasserman, K. Faust, D. Iacobucci, M. Granovetter,
  {\it Social Network Analysis: Methods and Applications} (Cambridge, 1994).
\bibitem{paul} G. Paul, S. Sreenivasan and H.E. Stanley, Phys. Rev. E
  \textbf{72}, 056130 (2005).
\bibitem{newman} M.E.J. Newman, Phys. Rev. Lett. \textbf{89}, 208701 (2002).
\bibitem{tanizawa} T. Tanizawa, G. Paul, R. Cohen, S. Havlin, and
  H.E. Stanley, Phys. Rev. E \textbf{71}, 047101 (2005).
\bibitem{pastor} R. Pastor-Satorras and A. Vespignani, Phys. Rev. E
  \textbf{65}, 036104 (2002).
\bibitem{holme} P. Holme, B.J. Kim, C.N. Yoon, and S.K. Han, Phys. Rev. E
  \textbf{65}, 056109 (2002).
\bibitem{newman0} M.E.J. Newman and M. Girvan, Phys. Rev. E \textbf{69},
  026113 (2004).
\bibitem{borgatti}S.P. Borgatti, Comp. \& Math. Org. Theory \textbf{12}, 21
  (2006).
\bibitem{footnote}Group of connected nodes known as ``component'' in the
  language of sociology.
\bibitem{11a} A. Bunde and S. Havlin, {\it Fractals and Disordered Systems}
  (Springer, 1995).
\bibitem{dietrich} D. Stauffer and A. Aharony, {\it Introduction to
  Percolation Theory} (Taylor \& Francis, London, 1994). 
\bibitem{cohen3} R. Cohen, et al., Phys. Rev. E \textbf{66}, 036113 (2002).
\bibitem{erdos} P. Erd\H{o}s and A. R\'{e}nyi, Publ. Math. (Debrecen)
  \textbf{6}, 290 (1959).
\bibitem{cohen4} R. Cohen, S. Havlin, and D. ben-Avraham, "Structural
  properties of scale free networks", Chap. 4 in {\it Handbook of graphs and
  networks}, Eds. S. Bornholdt and H. G. Schuster (Wiley-VCH, 2002).
\bibitem{newman2} M.E.J. Newman, SIAM Rev. \textbf{45}, 167 (2003).
\bibitem{scb} WWW.SCB.SE
\bibitem{viboud} C. Viboud, O.N. Bj\o rnstad, D.L. Smith, L. Simonsen, M.A.
  Miller, and B.T. Grenfell, Science \textbf{312}, 447 (2006).
\bibitem{gabor} G. Cs\'{a}nyi and B. Szendr\H{o}i, Phys. Rev. E \textbf{69},
  036131 (2004).
\bibitem{konstantin} K. Klemm and V.M. Egu\'{i}luz, Phys. Rev. E \textbf{65},
  057102 (2002).
\bibitem{shai} S. Carmi, S. Havlin, S. Kirkpatrick, Y. Shavitt and E. Shir,
  cond-mat/0601240 (2006).

\end{thebibliography}
\end{document}